\documentclass[journal,onecolumn,10pt]{IEEEtran}
\usepackage{graphicx}
\usepackage{amsmath}
\usepackage{amssymb}
\usepackage{tikz}
\usepackage[utf8]{inputenc}
\usepackage[T1]{fontenc}
\usepackage{lmodern}
\usepackage{float}
\usepackage{dblfloatfix} 
\usepackage{cite}
\usepackage{wrapfig} 
\usepackage{lscape}
\usepackage{multirow}
\usepackage{rotating}
\usepackage{setspace}
\usepackage{caption}
\usepackage{subcaption}
\ifCLASSINFOpdf
\else
\fi

\begin{document}
	
	\title{Plane Wave Dynamic Model of Electric Power Networks with High Shares of Inverter-Based Resources}

	\author{
		Amirhossein~Sajadi$^{1}$,~ 
		Bri-Mathias~Hodge$^{1,2,3}$
		
		\footnotesize{$^{1}$Renewable and Sustainable Energy Institute at the University of Colorado Boulder, 4001 Discovery Drive, Boulder, CO 80303, USA} 
		
		\footnotesize{$^{2}$Department of Electrical, Computer and Energy Engineering, University of Colorado Boulder, 425 UCB, Boulder, CO 80303, USA} 
		
		\footnotesize{$^{3}$Grid Planning and Analysis Center at the National Renewable Energy Laboratory (NREL), 15013 Denver W Pkwy, Golden, CO 80401, USA}

		\footnotesize{Corresponding author: Amir.Sajadi@colorado.edu}

	}

	\markboth{~}
	{~}
	
	\maketitle

	\begin{abstract} 
		Contemporary theories and models for electric power system stability are predicated on a widely held assumption that the mechanical inertia of the rotating mass of synchronous generators provides the sole contribution to stable and synchronized operation of this class of complex networks on subsecond timescales. Here we formulate the electromagnetic momentum of the field around the transmission lines that transports energy and present evidence from a real-world bulk power network that demonstrates its physical significance. We show the classical stability model for power networks that overlooks this property, known as the "\textit{swing equation}", may become inadequate to analyze systems with high shares of inverter-based resources, commonly known as "\textit{low-inertia power systems}". Subsequently, we introduce a plane wave dynamic model, consistent with the structural properties of emerging power systems with up to 100\% inverter-based resources, which identifies the concept of inertia in power grids as a time-varying component. We leverage our theory to discuss a number of open questions in the electric power industry. Most notably, we postulate that the changing nature of power networks with a preponderance of variable renewable energy power plants could strengthen power network stability in the future; a vision which is irreconcilable with the conventional theories.  
		
	\end{abstract}

	\IEEEpeerreviewmaketitle

	
	\section{Introduction}  
	
	An alternating current (AC) power grid is a complex network whose synchronization on a short timescale (subsecond to a few seconds), from the perspective of contemporary theories and models, is analogous to abundance of the mechanical inertial momentum provided by synchronous generators \cite{motter2013spontaneous,denholm2020inertia}. Also present in this network is the momentum of the electromagnetic field that transports the energy across its links which, despite being significant and fundamental to the AC power transmission, has never been studied. Such treatment has held valid for more than half a century since the advent of modern power grids \cite{liacco1967adaptive} mainly because of the structural properties of conventional power networks, especially synchronous machines, that allowed for a series of simplifications which result in treating this complex network as an electromechanical network rather than an electromagnetic network. Additionally, the classical formalism for power networks is a simplified approximation of the electromagnetic mechanism of energy flow, primarily developed for smaller networks (regional power grids), where the surface volume within which the carrier field is bounded were much smaller compared to the interconnections of today, with exceptionally large surface volume, that spread across continents. For example, the North American Eastern and Western Interconnections each cover broad swaths of the North American landmass and the synchronous grid of Continental Europe covers 24 countries. These premises have obscured a number of contemporary issues in power system dynamics, such as forced oscillations \cite{ye2016analysis,sarmadi2015inter} and grid strength \cite{dozein2021simultaneous, AEMO2020strength} until the recent wide-spread adoption of inverter-based renewable resources that break away from the synchronous generator-based paradigm. 
	
	The incorporation of the electromagnetic phenomena in power network analyses requires its treatment as electromagnetic waves, the type of fundamental physical forces that AC electricity generates for transfer of power, and their associated properties and attributes. Subsequently, the presentation of power network dynamics will become two dimensional to simultaneously describe the nodal frequency and voltage dynamics at the network level. This emerging direction of research has attracted many to conduct theoretical studies that are extensions to the classical model of frequency dynamics, known as the "swing equation" \cite{machowski2020power}, and has been spearheaded by the recently introduced concepts of \textit{complex-frequency} \cite{milano2021complex, milano2021geometrical, moutevelis2023taxonomy, he2022complex}, \textit{augmented synchronization} \cite{yang2021augmented}, and \textit{phasor isomorphism} \cite{li2023intrinsic}, all of which stem from a circuit analysis approach to address the dynamics of modern power networks. This work takes a different approach and builds upon the existing body of fundamental literature on the characterization of power networks and power flow from the perspective of electromagnetism \cite{olsen2015high,emanuel2011power,kirtley2020electric,bergen2009power}, which yields interesting new findings. 
	
	Here we advance the field by formulating the momentum stored in the electromagnetic field around the transmission lines in bulk power networks directly from Maxwell's laws of electromagnetism. We present evidence from phasor measurement units in a real-world bulk power network that demonstrate the physical significance of this component, whose contributions have thus far been overlooked in the electric power network literature. Subsequently, we develop a plane wave model for dynamics and stability in power systems based on the electromagnetic coupling of frequency and voltage, which fills a number of gaps in current power system dynamics theory and presents an elegant solution to characterize the dynamics and structural properties of the modern power grid. In contrast to the conventional understanding that considers the mechanical inertia of generators as invariant and as the sole contributing component to system inertia \cite{machowski2020power,sauer2017power}, our results demonstrate that the concept of effective inertia in power networks is time-variant and has contributions dependent both on the network conditions and the generator structural properties.
	Our plane wave dynamic model formalizes the theory of the structural strength of a power network and identifies the contributing factors, which enables the adoption of mechanisms for the enhancement of system strength. We find the necessary conditions under which the changing nature of power networks with a preponderance of variable renewable energy power plants can actually lead to more stable electric energy delivery in the future, a vision which is irreconcilable with the conventional theories and models of stability in power networks.

	Our study has broader impacts beyond power grids and advances the field of complex networks by offering direct observation into the interactive dynamics of harmonic oscillators that are coupled via an electromagnetic link at high power and extremely large scales, a configuration that is unlikely to be reproduced in a laboratory setting. This could help with further exploring the subject of stability and synchronization of large-scale electromechanical-electromagnetic coupled complex networks and with the observation, analysis, and understanding of these complex phenomena that occur in the real world.

	\section{Results}

	\subsection*{Electromagnetic Wave Propagation and Power Flow in Power Networks}
	
	The theory of power flow in electric power networks is explained by the Poynting theorem \cite{poynting1884xv}, an extension of Maxwell's equations \cite{maxwell1865viii}. Maxwell's equations explain that the electromagnetic field, where energy is stored, is composed of both electric and magnetic field vectors, $\vec{E}$ and $\vec{H}$, respectively, that are inseparable and lie in a plane which is transverse to the axis of energy propagation. The Poynting vector, $\vec{S}=\vec{E} \times \vec{H}$, which is a polarized wave, composed of mutually perpendicular waves differing in time and amplitudes \cite{collin1990field}, quantifies the amount of power that leaves the surface in which the energy is stored and is equivalent to the negative of the work done on the charges within the volume minus the losses. The Poynting theorem links Maxwell's laws of electromagnetism to Kirchhoff's laws of circuit theory, which form the basis for the power flow equation used in electric power networks \cite{emanuel2004poynting}. The Poynting Vector in the integral form is presented as \cite{balanis2012advanced,wolski2011theory}:
	\begin{align}
		\begin{split}
			\oint_\mathcal{A} \vec{S} \cdot d\vec{\mathcal{A}}  = 
			- \frac{\partial}{\partial t} \Big(
			\int_\mathcal{V_H} \dfrac{1}{2}  \cdot \mu \cdot \vec{H}^2   \cdot d\mathcal{V_H}
			+ \int_\mathcal{V_E} \dfrac{1}{2}  \cdot \varepsilon \cdot \vec{E}^2 \cdot d\mathcal{V_E}
			\Big)
			-  \int_{\mathcal{V}_\Omega} \vec{E} \cdot \vec{J} \cdot  d\mathcal{V}_\Omega
		\end{split}
	\end{align}
	where $\mathcal{A}$ is the cross-sectional surface area through which the electric power is transferred. $\mathcal{V_H}$ and $\mathcal{V_E}$ are the volumes of free space around the line where magnetic and electric fields are enclosed, respectively, and $\mathcal{V}_\Omega$ is that of the space within the conductor in which resistive losses occur. $\varepsilon$ and $\mu$ are the electric permittivity and magnetic permeability, respectively, the material characteristics of a conductive medium and define its response to electric and magnetic fields \cite{wolski2011theory}. This formulation suggests the electric power, denoted by $s=\oint_\mathcal{A} \vec{S} \cdot d\vec{\mathcal{A}}$, is the rate of change of energy stored in the electromagnetic field, $w$, with time $t$, $s=\frac{\partial }{\partial t}w$, that is the aggregation of the power that exits the surface area of a conductor and the power stored in the magnetic and electric fields, also known as total power. In an AC power network, which is a harmonic field, the energy function of the electromagnetic field can be calculated by:
	\begin{align}
		\begin{split}
			w & = 
			-  \dfrac{1}{2} \int_\mathcal{V_H}  \Big( \dfrac{\vec{\Phi}}{l \cdot  d}\cdot \dfrac{\vec{i}^*}{2 \cdot \pi \cdot  r_\Phi} \Big)  \cdot d\mathcal{V_H} 
			-  \dfrac{1}{2} \int_\mathcal{V_E}  \Big(  \dfrac{\vec{q}}{2 \cdot  \pi  \cdot  r_q \cdot l}  \cdot \dfrac{\vec{v}^*}{d} \Big) \cdot d\mathcal{V_E} 
			- \int \int_{\mathcal{V}_\Omega} \dfrac{R}{l \cdot \pi \cdot  \hat{r}^2} \cdot   \vec{i}^* \cdot   \vec{i} \cdot d\mathcal{V}_\Omega \cdot  dt \\
		\end{split}
	\end{align}
	where $\vec{i}$ is electric current in the conductor that connects nodes $i$ and $j$ and according to Faraday's law, $\vec{\Phi}=\int_{\mathcal{A}_{\Phi}} \mu \cdot \vec{H} \cdot d\mathcal{A}_{\Phi} =\int_{\mathcal{A}_{\Phi}} \vec{B} \cdot d\mathcal{A}_{\Phi}$, is the magnetic flux linkage that passes through the area $\mathcal{A}_{\Phi}$ of the open space between the conductors in overhead lines at point $i$. $\mathcal{A}_{\Phi}=l \cdot d$, and $\vec{B}$ is the magnetic flux density. $l$ is the length of the transmission line, $d$ is the distance between adjacent conductors of the line, $r_\Phi$ is the radius of the magnetic field lines as formed in the free space around the conductor. $\vec{v}$ is voltage at any point along the line and according to Gauss's law, $\vec{q}=\int_{\mathcal{A}_{q}} \varepsilon \cdot \vec{E} \cdot d\mathcal{A}_{q}= \int_{\mathcal{A}_{q}} \vec{D} \cdot d\mathcal{A}_{q}$, is the electric charge within the area $\mathcal{A}_{q}$ and $\vec{D}$ is the electric flux density, assuming electric charges are distributed uniformly on the conductor surface along the line, denoted by $\mathcal{A}_{q}$, and ignoring the ground capacitance. The area is described by $\mathcal{A}_{q}=2 \cdot \pi  \cdot r_q  \cdot l$ where $r_q$ is the radius of the electric field lines as formed in the free space around the conductor. $R= \frac{\rho l}{\pi \hat{r}^2}$ is electric resistance where  $\hat{r}$ is the radius of the conductor and $\rho$ is the electric resistivity of the conductor, defined as $\vec{E}=\rho  \cdot \vec{J}$, a material property that measures how strongly it resists electric current. 
	The total electric charges inside the area $\mathcal{A}_{q}$ can be expressed as $q=\oint_{\mathcal{A}_{q}} \vec{D} \cdot d\mathcal{A}_{q}=C\cdot \vec{v}$ where $C$ is the effective shunt capacitance and $\vec{v}$ is voltage. The magnetic flux linkage can be also expressed as $\Phi=\oint_{\mathcal{A}_{\Phi}} \vec{B} \cdot d\mathcal{A}_{\Phi}=L\cdot \vec{i}$ where $L$ is the effective series inductance. $^*$ is the conjugate operator. Replacing the magnetic flux linkage and electric charge values yields:
	\begin{align}
		\begin{split} 
			w&=-  \dfrac{1}{2}  \Big( L\cdot \vec{i} \cdot \vec{i}^* + C\cdot \vec{v} \cdot \vec{v}^*  \Big)  -   \int R  \cdot  \vec{i}^* \cdot   \vec{i}  \cdot  dt    
		\end{split}
	\end{align}
	This is the total field energy inside a volume $\mathcal{V}$. The power flow equation then can be obtained by simply taking the time derivative of the energy function, $w$, as $\frac{\partial  w}{\partial t} = -  L\cdot \vec{i}^* \cdot   \frac{\partial  }{\partial t} \vec{i}-   R  \cdot  \vec{i}^* \cdot   \vec{i} $, where the parasitic capacitance, $C$, is reasonably ignored. By defining the current in phasor representation as $\vec{i}= \overline{Y}_{ij} \cdot ( V_j \cdot e^{\boldsymbol{j} (\omega_j t +\delta_j)}-V_i \cdot e^{\boldsymbol{j} (\omega_i t +\delta_i)}) $ where $V_i$ and $V_j$ are the the root-mean-square (RMS) voltage amplitudes at the $i$ and $j$ nodes and $\delta_i$ and $\omega_i$ and $\delta_j$ and $\omega_j$ are the phase angle and frequency of the respective node. $\overline{Y}_{ij}$ is the admittance of the line that connects the two nodes in phasor notation. The power flow equation for the $i$th node, in quasi-steady state, $\omega_i \approx \omega_j$, consists of two terms, $\frac{\partial  w}{\partial t} = \frac{\partial  w_S}{\partial t}  + \frac{\partial  w_\Omega}{\partial t}$ and each term is given as:
	\begin{align}
		\begin{split} \label{eq:power_flow_eq}
			& \frac{\partial  w_S}{\partial t}=   L_{ij} \cdot \overline{Y}_{ij}  \cdot \overline{Y}_{ij}^* \cdot V_i \cdot  (V_j  \cdot  e^{\boldsymbol{j} (\delta_i-\delta_j)} - V_i )  \cdot  \dfrac{d  \delta_i  }{d t}   \\  
			& \frac{\partial  w_\Omega}{\partial t}   =  R_{ij}  \cdot \overline{Y}_{ij} \cdot \overline{Y}_{ij}^* \cdot V_i \cdot \Big( V_j   \cdot  e^{(\delta_j-\delta_i)} -  V_i \Big) \\  
		\end{split}
	\end{align} 
	
	It is commonly represented in phasor form as $ \frac{\partial  w}{\partial t} =   \overline{Y}_{ij}^* \cdot V_i \cdot  (V_j  \cdot  e^{\boldsymbol{j} (\delta_i-\delta_j)} - V_i ) $, which is the well-known power flow equation for power systems. 
	The first term, $\frac{\partial  w_S}{\partial t}$ describes the power that exits the surface area of the conductor as the delivered power.
	The second term, $\frac{\partial  w_\Omega}{\partial t} $ describes the dissipated power in the conductor as the ohmic losses.
	
	Equation \eqref{eq:power_flow_eq} is very important because it describes the propagation of energy across a power network through the electromagnetic field that is formed around the transmission lines and highlights the crucial contributions by the electric angle, denoted by $\delta_i$, and the voltage magnitude, denoted by $V_i$. The electric angle is directly linked to the strength of the magnetic field - the value of $(\delta_i-\delta_j)$ determines amplitude of the magnetic field along the path between $i$th and $j$th nodes, and $\frac{d \delta_i}{dt} = \omega_i$ determines the frequency of its oscillations. The voltage magnitude is also directly linked to the strength of the electric field - the value of $V_i$ determines amplitude of electric the field at $i$th node, and the frequency of its oscillations coincides with that of the magnetic field.

	\subsection*{Electromagnetic Momentum in High Voltage Transmission Networks}
	
	An electromagnetic field around a transmission line, in addition to transferring power, also stores momentum as predicted by Maxwell \cite{maxwell1865viii} and later described by Thomson \cite{thomson284elements}. 
	This momentum stems from the mass equivalent of the energy flow in the electromagnetic field around the transmission line \cite{johnson1994electromagnetic}, which is a fundamental property of an electromagnetic field and must be included in the analyses according to the law of conservation of momentum \cite{bak1994energy}. Most commonly, AC electricity propagates in the form of an elliptical polarized wave, where the electric and magnetic fields differ in amplitude and phase, yet both oscillate at the same frequency. The amount of power transferred in an electromagnetic field is determined by the amplitude of each field, whereas the phase difference between the two components is directly defined by the power factor for alternating current power systems. The power factor is a ratio of the real part of the Poynting Vector (active power) to its absolute value (total -- or apparent -- power, which includes reactive power, i.e. the imaginary part of the Poynting Vector). The power factor determines the nature of the power transferred - capacitive, inductive, or resistive, therefore the nature of polarization, whether linear or elliptical is determined. If the polarization is elliptical, then its direction, clockwise or counterclockwise, is determined by whether it has a leading or lagging power factor. Most common operating conditions in power grids are at high power factor, with active power transfer dominating, therefore the polarization is close to the linear form. Accordingly, the line momentum, denoted by $M_{l}$, can be calculated using the Poynting vector, $\vec{S}$, and integrating the momentum density over the enclosed space:
	
	\begin{align}
		\begin{split} \label{eq:line_momentum}
			M_{l} &= \int_\mathcal{V}  \frac{1}{c^2} \cdot  \vec{S} \cdot  d\mathcal{V}\\
		\end{split}
	\end{align} 
	where $c$ is the speed of light, $\vec{S}$ is the Poynting Vector, and $\mathcal{V}$ is the volume of the surface bounding the electromagnetic field around the conductor throughout the entire length of the line. The symbol $\cdot$ is the dot product operator. Eq. \eqref{eq:line_momentum} makes it evident that the amount of electromagnetic momentum is directly proportional to the loading level of the lines which determines $\vec{S}$, and their length which determines the volume $\mathcal{V}$. 
	A change in the flow of power in a power line changes the energy density in its electromagnetic field which is directly proportional to the amplitudes of its constituting waves. Therefore, the most prominent impact of the line electromagnetic momentum in an electric power network is experienced by its electric angles.
	
	We wish to highlight two additional noteworthy points. First, the electric circuit of a transmission system contains a large number of transformers that are highly magnetic components. However, their electromagnetic field is locally bounded to the location of their operation, yielding a small volume of enclosed space. Consequently, their independent electromagnetic momentum is relatively small, but because their electromagnetic link is a part of the overall Poynting vector that transfers power between two nodes, their impact is aggregated in our analysis. Second, the bundled structure of conductors in transmission lines may influence the cross-sectional surface $\mathcal{A}$, but the Poynting vector $\vec{S}$ is the power density thus when integrated over the enclosed space volume $\mathcal{V}$, it only becomes a function of the net amount of power transferred and the length of the transfer, independent of the cross-sectional surface area. Therefore, this formulation is independent of the nuances of line structural design.

	\subsection*{Plane Wave Dynamic Model for Power Networks}
	
	The impetus behind the dynamic analysis of a power network is to understand and capture the transients that may appear following a disturbance along the electromagnetic link established between its generators and loads via the interconnecting transmission lines. The power flow equations, shown in Eq. 
	\eqref{eq:power_flow_eq}, attested to a direct dependence of energy propagation across a power network to the electric angle and voltage magnitude of all of its nodes. 
	Accordingly, we derive the plane wave dynamic equations to describe the electric angle and voltage at the $i$th node of a power network, denoted by $\delta_i$ and $V_i$, respectively, as: 
	\begin{align}
		\begin{split}
			\ddot{\delta}_i & = \mathbf{M}_i^{-1} \Big( P_{g_i} - D_i \cdot (\omega_i-\omega_s) - f_i(\delta , V) \Big) \\
			\dot{V}_i & = T_{v_i}^{-1} \Big( E_{g_i} -  V_i   - g_i(\delta , V)  \Big)
		\end{split}
	\end{align} 
	where 
	$\mathbf{M_i}=M_{g_i}+M_{l_i}$ is the total momentum at $i$th node which is the summation of the momentum of the generator connected to $i$th node, denoted by $M_{g_i}$, and the momentum associated with the lines connected to $i$th node, denoted by $M_{l_i}$. The line momentum is the aggregate of the momenta values of all $j$ number of lines that are directly connected to the $i$ node; $M_{l_i}=\sum_{k=1}^{j} M_{l_{ik}}$ where $M_{l_{ik}} \neq M_{l_{ki}}$ since a line momentum is conditioned on the direction of flow of power.  
	$T_{v_i}$ the transient voltage time constant of $i$th generator that articulates the number of seconds it takes for its terminal voltage to change.
	$D_i$ is frequency damping functions that approximate the frequency response/control actions, respectively, and hold distinct values for each generation technology. 
	This formulation allows us to consider all three power conversion technologies that are in the landscape of the transition to 100\% renewable-based power networks. The first technology is the synchronous generator; an overwhelming majority of existing generation facilities are equipped with this technology, including hydro, nuclear, natural gas, and coal-fueled power plants \cite{machowski2020power}. The second and third technologies are based on power electronic inverters (henceforth referred to as inverter-based resources (IBRs)) which interface variable renewable energy sources and energy storage units with the grid. They can be categorized as (2) grid-following inverters (referred to as GFL, henceforth) and (3) grid-forming inverters (referred to as GFM, henceforth), including all three GFM categories of virtual synchronous machine (VSM) \cite{beck2007virtual}, multi-loop droop GFM (GFM-droop) \cite{lasseter2019grid}, and virtual oscillator control (VOC) \cite{sinha2015uncovering} devices. Each technology can be modeled by distinct momentum values $M_{g_i}$ and $T_{v_i}$ and damping function, $D_{i}$. 
	Network flow function are: $f_i(\delta , V)= V_i ^2 \cdot  Y_{ii}  \cdot cos(\alpha_{ii})  + \sum_{\substack{j=1 \\ i\neq j}}^{n}  V_i \cdot  V_j \cdot Y_{ij} \cdot cos(\delta_i-\delta_j-\alpha_{ij})$ for real power flow and 
	$g_i(\delta , V) =|Z_m \cdot  \big(  V_i \cdot Y_{ii} \cdot e^{\boldsymbol{j}(\delta_i+\alpha_{ii})} + \sum_{\substack{j=1 \\ i\neq j}}^{n}  V_j \cdot Y_{ij} \cdot e^{\boldsymbol{j}(\delta_j+\alpha_{ij})}\big)| $ for voltage, 
	where $\overline{Y}_{ij}$ is the mutual admittance between nodes $i$ and $j$, also expressed in complex form as $\overline{Y}_{ij}=Y_{ij} \angle \alpha_{ij}$, and $\overline{Y}_{ii}$ is the self admittance of $i$th node. 
	$Z_m$ is the magnetizing impedance of $i$th generator; In synchronous generator, that is the \textit{armature reaction reactance} or \textit{magnetizing reactance} and in inverter-based generators, that is the impedance between the LCL/LC filter input and its capacitor.
	$P_{g_i}$ is the $i$th generator's real power that is converted from the primary energy source into AC electric power by the interfacing device (e.g synchronous machine or inverter). The $E_{g_i}$ is the $i$th generator's excitation voltage; in synchronous generator, it is known as the \textit{excitation emf} and in inverter-based generators that are equipped with voltage source converters (VSC), this is the AC side of the semiconductor switches that is the input to the LCL/LC filter.
	$\omega_i=\frac{d\delta_i}{dt}$ is the frequency at $i$th node and $\omega_s$ is the network synchronization frequency. 
	Its fixed points form the necessary condition of power balance in the power network: active power produced should be equal to the power consumed and excitation voltage should be such to regulate voltage magnitude within the standard operational range at all times, which is intuitive.

	Our formulation explicitly establishes the link between nodal frequency and voltage by the direct incorporation of differential equations that describe their dynamics at the network-level. This characterization is consistent with the recently established concept of complex frequency \cite{milano2021complex} that uses the localized curvature of the frequency-power manifold to establish a direct link between the azimuthal and radial components, with the assumption that the relationship between the frequency changes in any two adjacent nodes in a power network of interest are linear \cite{milano2021geometrical}. Thus, our findings contribute to the existing body of literature on the two dimensional nature of synchronization in power networks and is corroborated by them \cite{milano2021complex, milano2021geometrical, moutevelis2023taxonomy, he2022complex, yang2021augmented,li2023intrinsic}, with the novel addition of the consideration of electromagnetic momentum stored in the field around the lines, which is an important distinction in our model. Modeling dynamics of power networks in two-dimensional space is a new direction of research and is a departure from the frequency stability convention, which is centered on a swing equation that describes the synchronization dynamics of power networks in a single-dimensional differential equation. The classical swing equation has served as the ground truth for studying various aspects of power system dynamics and synchronization for more than half a century, accompanied by a series of algebraic equations to estimate network-level voltage. Our plane wave dynamic model presents a similarities to the classical swing equation, because it is a more complete form of swing equation. In our formulation, if we follow the traditional assumptions, the equation reduces to the classical swing equation. If we, (1) ignore the term related to the electromagnetic momentum, $M_{l_i}$, and (2) assume the reactive power support of the generator is so generous that the availability of required reactive power, and subsequently voltage regulation, is a given, i.e. $E_{g_i} = V_i + g_i(\delta , V)$ and the dynamic voltage differential terms become algebraic equality terms, this formalism reduces to $M_{g_i}  \cdot \ddot{\delta}_i  = P_{g_i}  -  D_{\omega_i} \cdot  (\omega_i-\omega_s) -  P_e $ where $P_e$ is the electric power that $i$th generator exports; that is exactly the well-known classical swing equation \cite{machowski2020power}.

	\subsection*{Numerical Validation: Plane Wave Model vs. Swing Equation}

	\begin{figure}[!b]
		\centering  
		\begin{subfigure}[b]{0.4\textwidth}
			\centering
			\includegraphics[width=.95\textwidth]{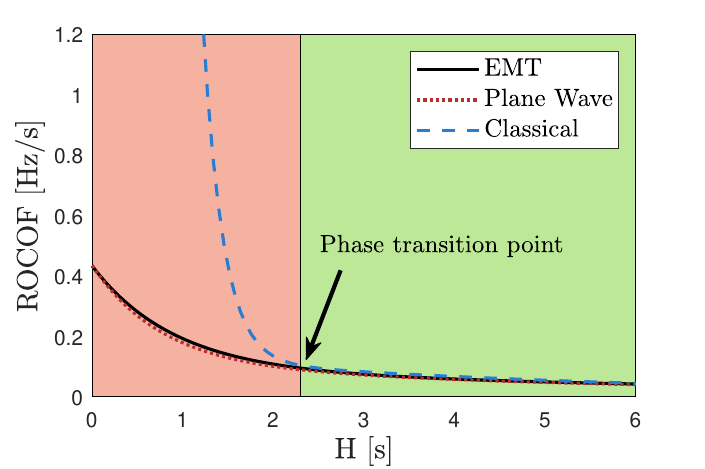}
			\caption{~}
			\label{fig:9-small}
		\end{subfigure}  
		\begin{subfigure}[b]{0.4\textwidth}
			\centering
			\includegraphics[width=.95\textwidth]{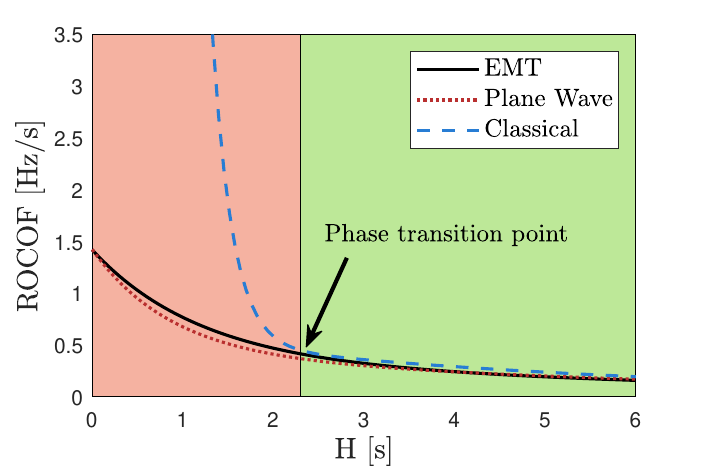}
			\caption{~}
			\label{fig:39-small}
		\end{subfigure}   
		\begin{subfigure}[b]{0.4\textwidth}
			\centering
			\includegraphics[width=.95\textwidth]{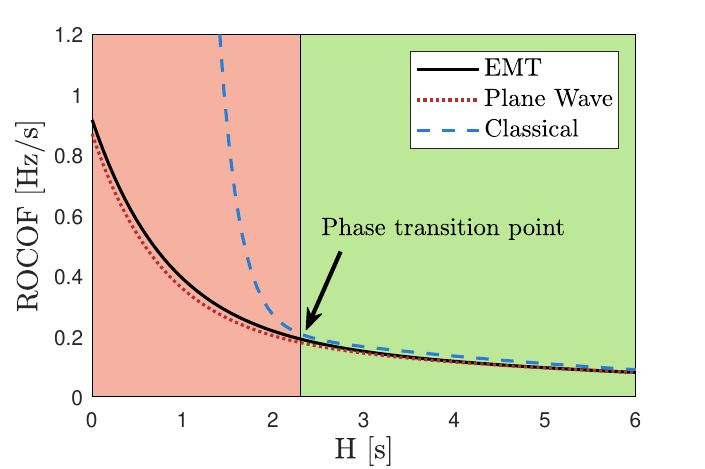}
			\caption{~}
			\label{fig:9-large}
		\end{subfigure}  
		\begin{subfigure}[b]{0.4\textwidth}
			\centering
			\includegraphics[width=.95\textwidth]{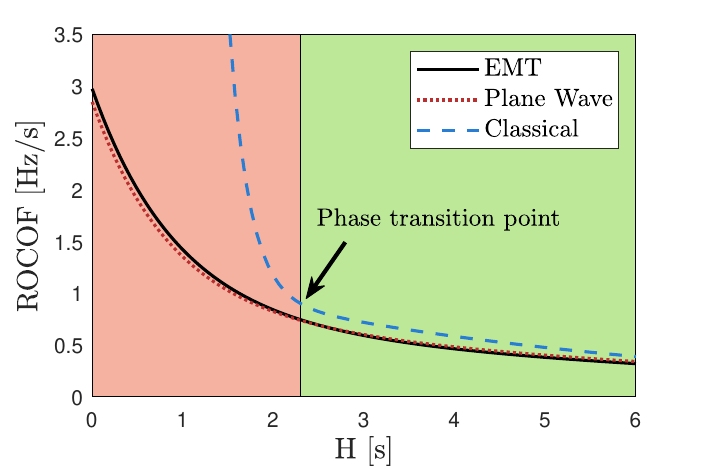}
			\caption{~}
			\label{fig:39-large}
		\end{subfigure}  
		\begin{subfigure}[b]{0.4\textwidth}
			\centering
			\includegraphics[width=.95\textwidth]{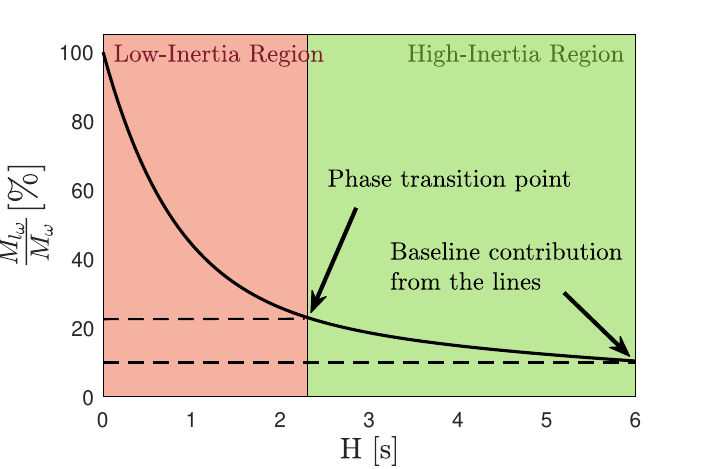}
			\caption{~}
			\label{fig:9-Ms}
		\end{subfigure}   
		\begin{subfigure}[b]{0.4\textwidth}
			\centering
			\includegraphics[width=.95\textwidth]{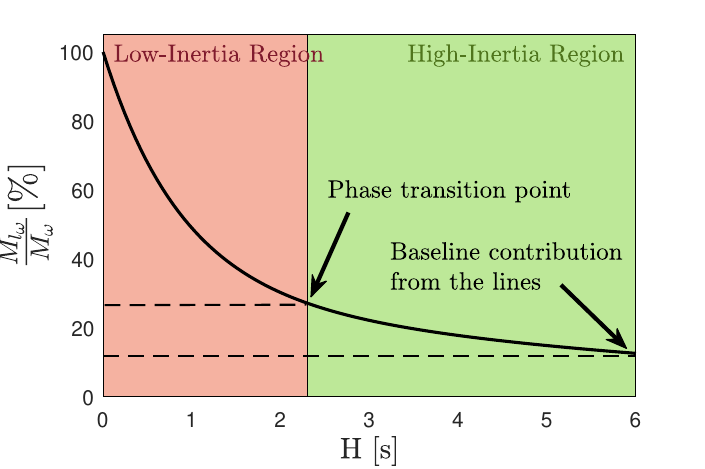}
			\caption{~}
			\label{fig:39-Ms}
		\end{subfigure}  
		\caption{\small{Assessment of power system dynamics with and without electromagnetic momentum. The results here signify the importance of the electromagnetic momentum of power lines in power system dynamics, especially during operating conditions with a preponderance of IBRs, where the momentum from generators is low (low-inertia systems - highlighted by the red zone).
		}}  
		\label{fig:PSCAD_9_39}
	\end{figure}

	To numerically validate our model and demonstrate the pivotal importance of the inclusion of the electromagnetic momentum in grid stability analyses, a comparative numerical analysis of our plane wave dynamics model with the numerical results of the classical model and an electromagnetic transient (EMT) simulation from Power Systems Computer Aided Design (PSCAD) - a renowned simulator for high-fidelity point on wave (POW) simulations - is shown in Fig. \ref{fig:PSCAD_9_39}. The test cases are two standard power system benchmarks: (1) the Western System Coordinating Council (WSCC) 9-node, 3-generator system \cite{sauer2017power} and (2) the New England (NE) 39-node, 10-generator system \cite{athay1979practical}. To emphasize the impact of the electromagnetic momentum of the field around transmission lines, we narrow down our focus to the system frequency dynamics. Following a disturbance, in the form of a load switch, we compute the rate of change of frequency (ROCOF) with our plane wave model and with the classical swing equation, and compare them with that of the PSCAD simulation. The results here show the system ROCOF profile with respect to a homogeneous inertia constant for generators, $H$. Momentum, $M$ and the inertia constant $H$ are related as $M=\frac{2H}{S}$ with $S$ being the generator rated capacity. In these plots, the green regions are "high inertia" operating conditions and red regions are "low inertia" operating conditions. 
	Figs. (\textbf{\ref{fig:9-small}}) and (\textbf{\ref{fig:9-large}}) show the 9-node WSCC system response to 4.5MW and 9MW disturbances, respectively. Figs. (\textbf{\ref{fig:39-small}}) and (\textbf{\ref{fig:39-large}}) show the 39-node NE system response to 307MW and 615MW disturbances, respectively. Figs. (\textbf{\ref{fig:9-Ms}}) and (\textbf{\ref{fig:39-Ms}}) show the percentage ratio of the electromagnetic momentum provided by the lines, $M_{l}$ - estimated by our theoretical analysis - to the total effective momentum of the system which is the sum of the momentum from generators and the lines, all with respect to changes of frequency, for the WSCC and NE systems, respectively. 
	
	We incrementally decreased the mechanical momentum provided by the generators to arrive at a case with only GFM generators that have insignificant practical momentum, thus all momentum in the system is provided by the electromagnetic field around the transmission lines.  We then conducted a simple model fitting using the sampled data points to create a continuous model. As is evident in Figs. (\textbf{\ref{fig:9-small}}), (\textbf{\ref{fig:9-large}}), (\textbf{\ref{fig:39-small}}), and (\textbf{\ref{fig:39-large}}) the traces of our plane wave model (depicted in dotted red lines) nearly perfectly match those of the EMT simulations (depicted in solid black lines). Whereas the traces of the classical swing equation (depicted in dashed blue lines) increasingly diverge from the EMT simulations as the generator inertia decreases. Results in Figs. (\textbf{\ref{fig:9-Ms}}) and (\textbf{\ref{fig:39-Ms}}) indicate the share of electromagnetic momentum as a percentage of the total effective momentum of the system, which monotonically increases with the reduction in the mechanical momentum provided by generators, reaching a nearly 100\% contribution in the all-GFM case, i.e. the $H\approx0$ data points in these plots. This signifies the role of electromagnetic momentum of power lines in power system dynamics, especially during operating conditions with a preponderance of IBRs, where the momentum from generators is low (low-inertia systems). We recognize that frequency response of GFMs, and subsequently their ROCOF, is a function of their control apparatus, specifically the droop value. For the case with the lowest inertia, all-GFM case, we employed droop coefficient of $5\%$ where a response is expected from the GFMs within a few milliseconds of control delay, just long enough to measure the ROCOF as the frequency deviates momentarily before the control response comes into effect. That being said, the inertial momenta of the lines observed in the GFM case were consistent with their values observed in the SG cases for all levels of machine inertia, for both systems, which indicates the measured values of momenta here were not distorted by the GFM control. Additionally, no load inertia was present in these experiments, therefore all momenta in the system was provided by either the generators or the transmission lines.

	The results here impugn the adequacy of the classical swing equation, for studying frequency stability in modern power networks, especially those with a preponderance of IBRs during low-inertia operating conditions, because the electromagnetic momentum becomes more prevalent and pivotal to consider, as evident in Figs. (\textbf{\ref{fig:9-Ms}}) and (\textbf{\ref{fig:39-Ms}}). The electromagnetic momentum has always been present in bulk high voltage transmission networks, but it has not been adopted in the modeling assumptions because the large momentum provided by the rotating masses of large synchronous machines, historically the nearly universal technology to generate electricity, dominated the network dynamics. In practical terms, disregard of electromagnetic momentum has gone unnoticed until now primarily because in most common operating conditions, i.e. power grids operated with high inertia, the error caused remains insignificant - the green region in Figs. (\textbf{\ref{fig:9-small}}), (\textbf{\ref{fig:9-large}}), (\textbf{\ref{fig:39-small}}), and (\textbf{\ref{fig:39-large}}). The evolution of the power industry has led to the contemporary theories of power system stability being saturated by the idea that the momentum is primarily contributed by the mechanical momentum of synchronous generators, and is therefore held as a static value \cite{machowski2020power,sauer2017power}. We depart from the convention and instead theorize that the effective inertial momentum of a power system is determined by a combination of transmission lines characteristics and the generator's properties and capacities and therefore a time-varying value. Subsequently, we define "low-inertia" operating condition where system dynamics cease to follow the classical swing equation and instead become heavily reliant on the network electromagnetic momenta - the red region in Figs. \ref{fig:9-small}, \ref{fig:9-large}, \ref{fig:9-small}, and \ref{fig:39-large}. We estimate a phase transition point at around $H=2.15$s where the mechanical momentum is at its critical value. If the mechanical momentum declines further from this value, then the dominant dynamics of the system change in nature and their governance principles switch from the laws of electromechanical forces to the laws of electromagnetic forces. Subsequently, the system departs the high-inertia region and enters the low-inertia region. At the phase transition point, the contribution of electromagnetic momentum to the overall momentum present in the system reaches $22.95\%$ in the 9-bus system and $27.1\%$ in the 39-bus system. These results suggest the existence of a baseline $10.2\%$ and $12.6\%$ contribution from the electromagnetic field to the overall system momentum during operational cases with the highest inertia from synchronous generators, $H=6$s. The analysis here confirms the accuracy of our plane wave dynamic model to precisely describe power system behavior for all operation conditions including high-inertia and low-inertia cases.

	\subsection*{Observation of Electromagnetic Momentum in Real-World Power Grids}

	\begin{figure}[h]
		\centering  
		\begin{subfigure}[b]{.83\textwidth}
			\centering		
			\includegraphics[width=.71\textwidth]{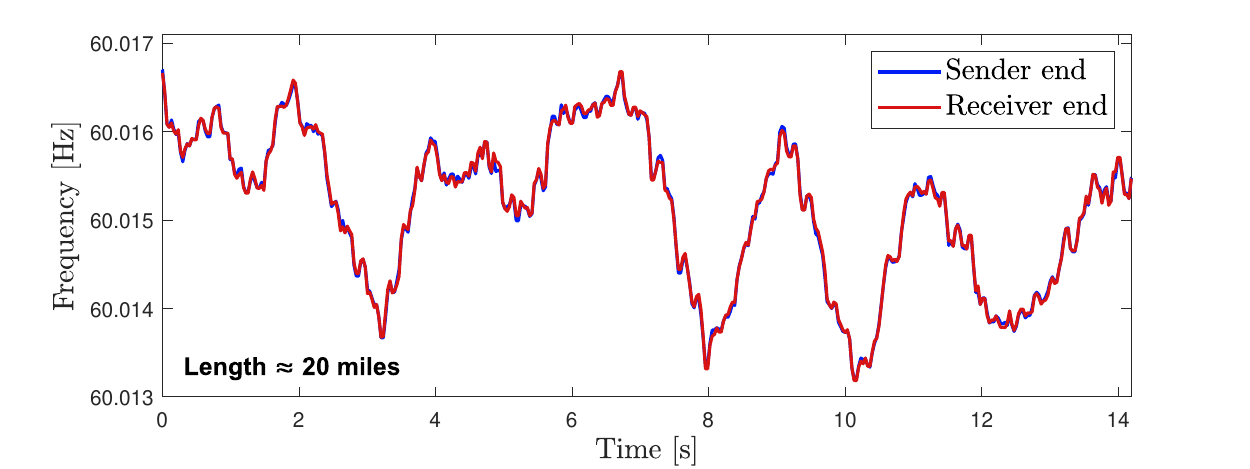}
			\includegraphics[width=.27\textwidth]{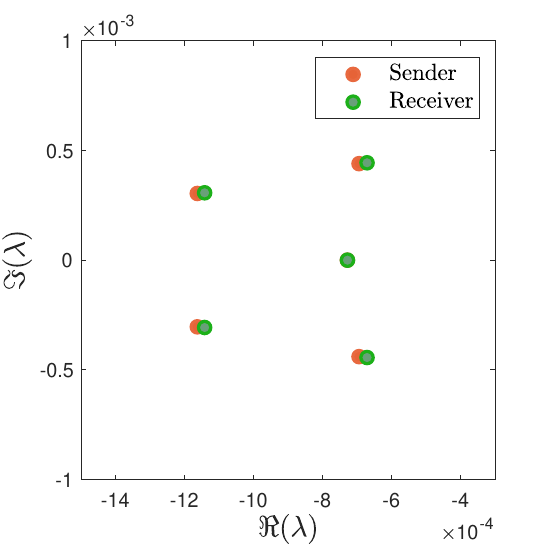}
			\caption{~}
			\label{fig:line1}
		\end{subfigure}    
		\begin{subfigure}[b]{.83\textwidth}
			\centering
			\includegraphics[width=.71\textwidth]{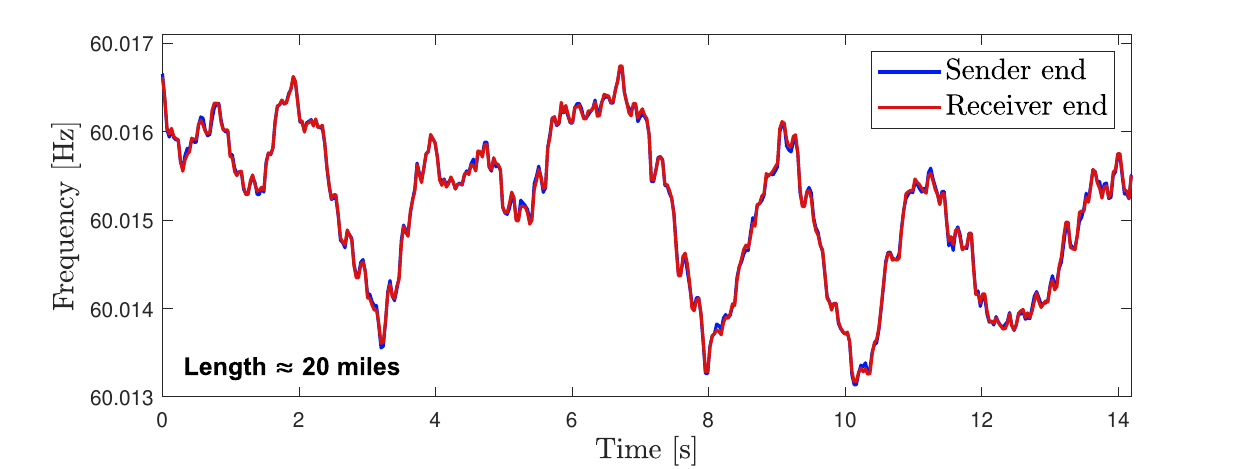}
			\includegraphics[width=.27\textwidth]{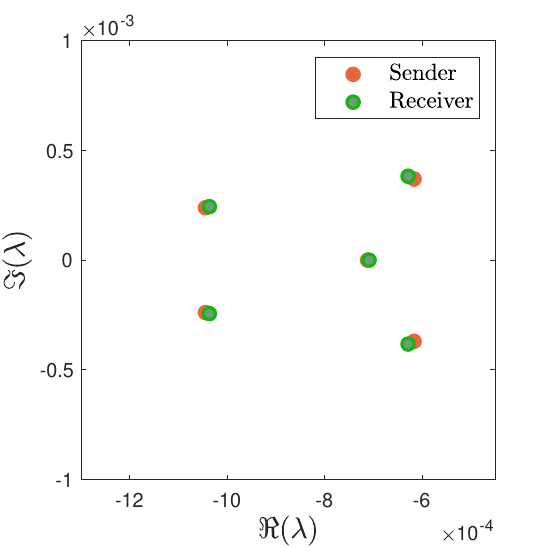}
			\caption{~}
			\label{fig:line2}
		\end{subfigure} 
		\begin{subfigure}[b]{.83\textwidth}
			\centering
			\includegraphics[width=.71\textwidth]{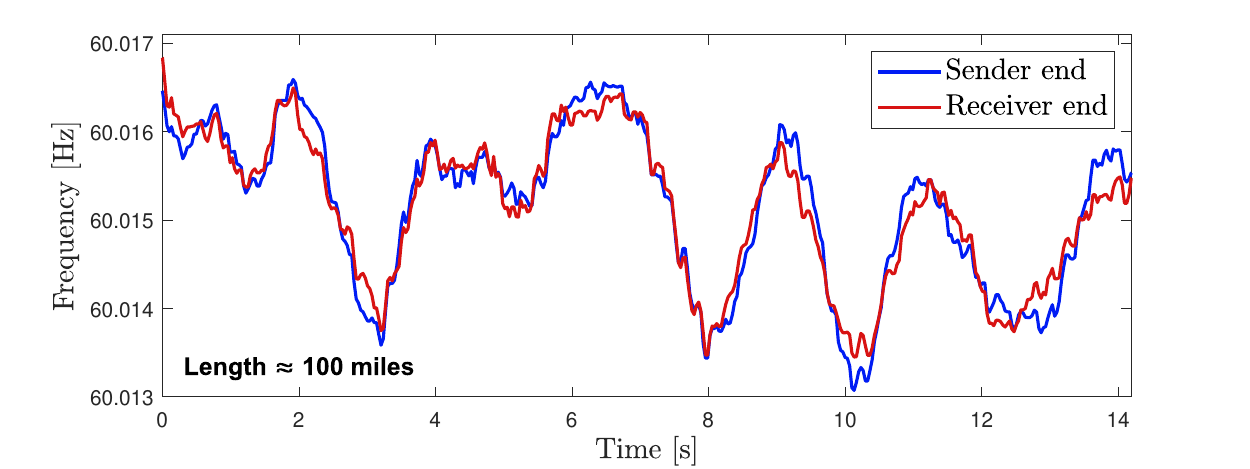}
			\includegraphics[width=.27\textwidth]{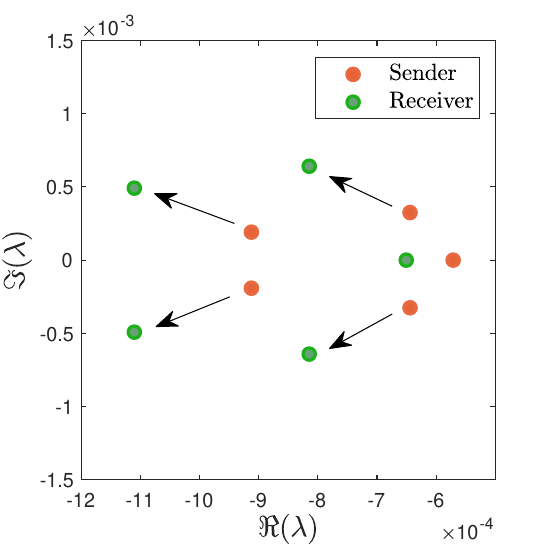}
			\caption{~}
			\label{fig:line3}
		\end{subfigure} 
		\begin{subfigure}[b]{.83\textwidth}
			\centering		
			\includegraphics[width=.71\textwidth]{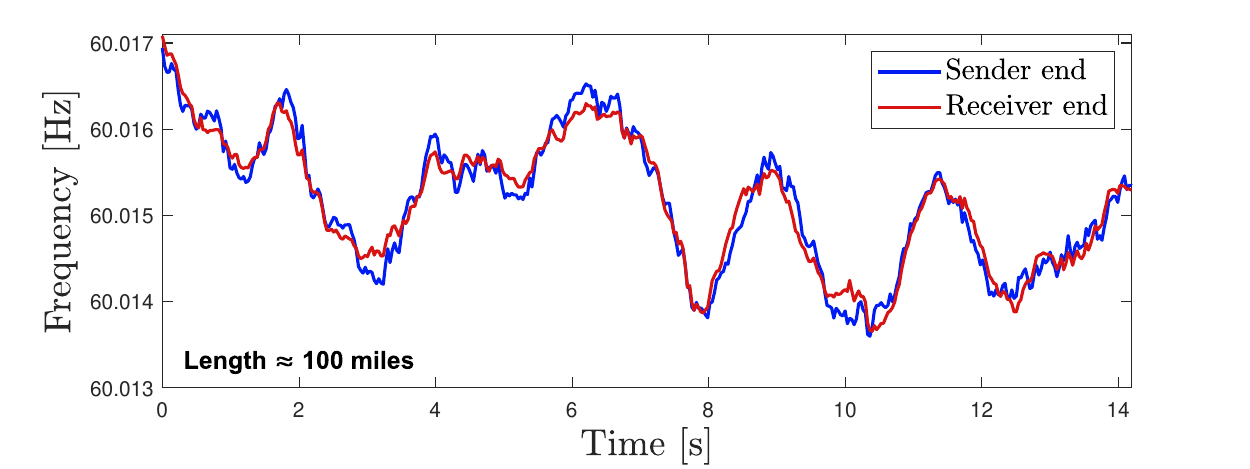}
			\includegraphics[width=.27\textwidth]{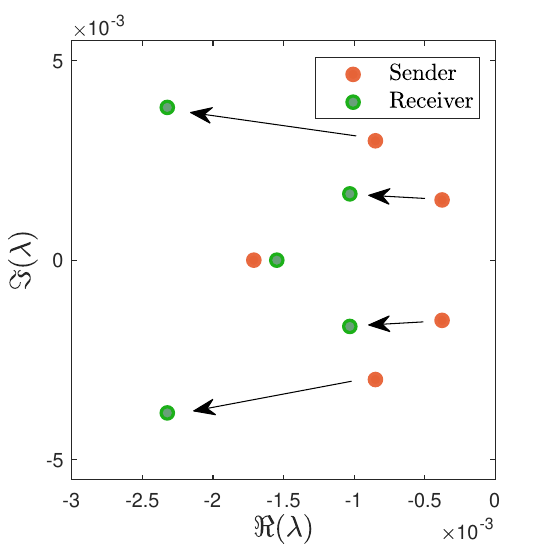}
			\caption{~}
			\label{fig:line4}
		\end{subfigure} 	
		\caption{\small{Observation of electromagnetic momentum in real-world high voltage power lines. The PMU data analysis indicates the existence of a considerable electromagnetic momentum in the longer lines and the significant impact this momentum could have on power system dynamics. The left column shows the time-domain trace, and the right column shows the migration of eigenvalues in complex plane. 
				%
		}}  
		\label{fig:SPP_data}
	\end{figure}
	
	To demonstrate the physical significance of the momentum of the electromagnetic field in the real-world, we were able to obtain measurement data recorded by phasor measurement units (PMUs) from the Southwest Power Pool (SPP), a major regional transmission organization (RTO) in North America that manages the electric grid for the central United States, covering 14 states. The measurement data that we acquired from different geographical locations are 
	synchronized by Global Positioning System (GPS). The analysis of this data provides empirical evidence of the influence of electromagnetic momentum in high-voltage transmission lines, as shown in Fig. \ref{fig:SPP_data}. We analyzed PMU data from the granular measurement of two ends of four different $345$kV transmission lines within the SPP power network, denoted $L_{i}, i=1,\cdots,4$. In all four lines, the cross-sectional surface, $\mathcal{A}$, through which power is transferred by the Poynting Vector, $\vec{S}$, is similar because they have similar voltage ratings and they are similarly loaded. They are distinct in terms of $\mathcal{V}$; the volume of the surface bounding the electric charges in their electromagnetic field because of different lengths; ${L_1}={L_2}\approx 20 \text{mi}$, making them relatively small and ${L_3} \approx {L_4}\approx 100 \text{mi}$, making them relatively large.  
	Figs. (\textbf{\ref{fig:line1}}) through (\textbf{\ref{fig:line4}}) show the frequency traces in the time-domain in the left column and the loci of migration of eigenvalues in the complex plane in the right column, for the two ends of the lines $L_{1}$ through $L_{4}$, respectively. 
	It is evident that the momentum stored in the electromagnetic field of the lines is proportional to their length which directly links the amount of this quantity to the surface volume that bounds the energy carrier field; the longer lines  where $\mathcal{V}$ is larger (at the same voltage level) store larger amounts of momentum that exhibit a greater capability in opposing the change of frequency. The frequency fluctuations in the receiving end are consistently smoothed out relative to those of the sending end - visible frequency differences between the two ends in Figs. (\textbf{\ref{fig:line3}}) and (\textbf{\ref{fig:line4}}) for lines 3 and 4 that are much longer than lines 1 and 2.
	
	It is worth noting that an electric charge that oscillates in the sending node of a transmission line experiences higher effective momentum (and subsequently higher inertia) than an electric change that oscillates in its receiving node. This is because as the energy flows along the line, the momentum becomes stored in the electromagnetic field that transports it by changing the internal structure of the median. Therefore, the amount of momentum delivered to the receiver ends becomes smaller and, consequently, the behavior of an electric charge at the receiving node is consistent with the dynamics of reduced inertia, which results in increased values of both the real and the imaginary terms of the complex eigenvalues and their migration further away from the origin to the left-hand side. A data-driven eigenstructure analysis of this data, using Prony least squares (PLS) method \cite{fernandez2018coding} confirms the existence of this mechanism in the PMU data we have obtained. In the complex plane, in the longer lines, the complex eigenvalues migrate further away from the origin in the receiver node relative to the sender node, Fig.(\textbf{\ref{fig:line3}}) and Fig.(\textbf{\ref{fig:line4}}). This migration is consistent with the signature of reduced inertia which indicates momentum is stored in the electromagnetic field around transmission lines as the energy is transferred (see \cite{sajadi2022synchronization} - Fig. 3(a)-1 reduced inertia case). 
	In the smaller lines, line 1 and 2, where $\mathcal{V}$ is smaller, the electromagnetic momentum is smaller, thus the frequency traces from the two ends of the lines are nearly identical, magnified in (\textbf{\ref{fig:line1}}) and (\textbf{\ref{fig:line2}}), and in the complex plane, all eigenvalues remained stationary. 
	This observation indicates the existence of a considerable electromagnetic momentum in lines 3 and 4 and the significant impact this momentum could have on power system dynamics. 
	The daily profiles of these lines indicate the time-varying nature of effective electromagnetic momentum throughout the day, as it is linked to the amount of energy transferred, and thus the different dispatch conditions throughout the day. The results shown here are only for single transmission lines and may seem very small by themselves, but keeping in mind the large number of such lines in an interconnection, it is clear that they can have a broader impact on power system dynamics. Our analysis here establishes the critical need to take the electromagnetic momentum of the field around high voltage transmission lines into account when modeling bulk power system dynamics with high shares of IBRs.

	\subsection*{Numerical Results for Frequency and Voltage Dynamics}
	
\begin{figure}[!]
		\centering  
		\begin{subfigure}[b]{.33\textwidth}
			\includegraphics[width=\textwidth]{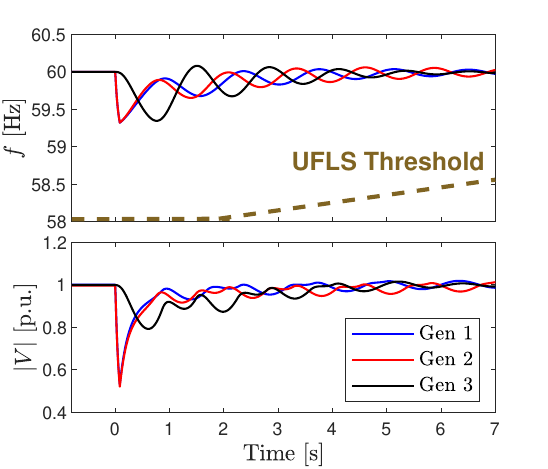} 
			\caption{~}  
			\label{fig:3SG2}    
		\end{subfigure}  
		\begin{subfigure}[b]{.33\textwidth} 
			\includegraphics[width=\textwidth]{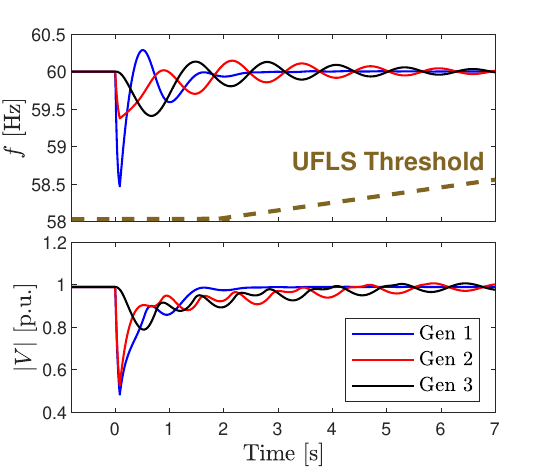}
			\caption{~}
			\label{fig:2SG-1GFM2}
		\end{subfigure}  
		\begin{subfigure}[b]{.33\textwidth} \label{fig:3GFM}
			\includegraphics[width=\textwidth]{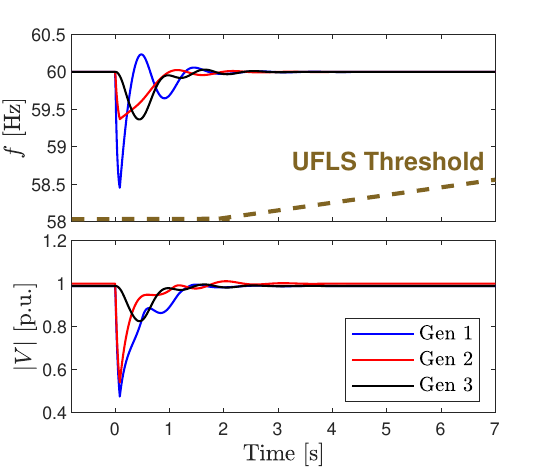}  
			\caption{~}  
			\label{fig:1SG-2GFM2}    
		\end{subfigure}  
		\begin{subfigure}[b]{.33\textwidth}     
			\includegraphics[width=\textwidth]{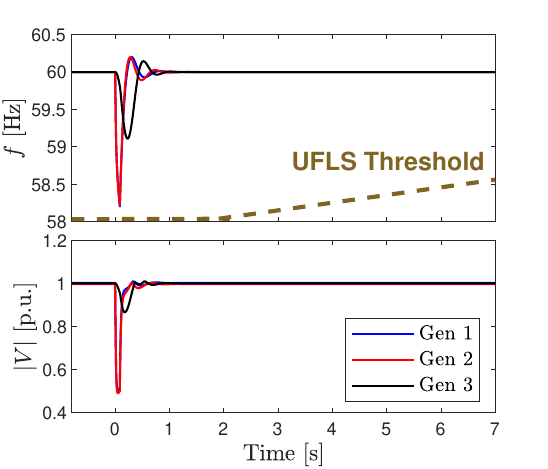}    
			\caption{~} 
			\label{fig:3GFM2}
		\end{subfigure}   
		\caption{\small{Dynamic stability in power network using the plane wave model. The results indicate an increase in the frequency deviations as the number of GFMs in the system increases whilst the SGs are present, but the transients could be damped faster. Nonetheless, the fast responding capability of GFM helps with the prevention of undesired activation of the under-frequency load shedding (UFLS) frequency relay - the UFLS threshold shown here is the continent-wide requirement for the North American power grid, according to the PRC-NPCC-02 standard \cite{NERC_UFLS_standard}.
		}}  
		\label{fig:MATLAB_9}
	\end{figure}

	To demonstrate the efficacy of our formulation in delineating the plane wave dynamics of power networks for the simultaneous consideration of frequency and voltage dynamics with any mixture of conventional synchronous generators and IBRs in the generation mix, the numerical results for the 9-node, 3-generator power system benchmark \cite{sauer2017power}, are shown in Fig \ref{fig:MATLAB_9}. This is a modified 9-bus system to create four different scenarios each with generation technology mixtures with a distinct combination of dynamical characteristics; (\textbf{\ref{fig:3SG2}}) 3 SGs, (\textbf{\ref{fig:2SG-1GFM2}}) 2 SGs and 1 GFM, (\textbf{\ref{fig:1SG-2GFM2}}) 1 SG and 2 GFMs, and (\textbf{\ref{fig:3GFM2}}) 3 GFMs - all GFMs are of the droop class. The perturbation is applied in the form of a three-phase symmetrical fault between generators 1 and 2 that is sustained for $83$ milliseconds, equal to 5 electrical cycles, followed by its clearance and system restoration to its pre-fault topology. In all four scenarios, the system loading is identical and the GFMs considered are of the multi-loop droop control variety. The results obtained here can be mapped to the VOC-GFM since the principle droop control is embedded in the VOC \cite{sinha2015uncovering} and the results from SG cases could be extended to VSM-GFM, given the nature of VSM-GFM is to emulate a SG. 

	The results show the time-domain traces of frequency and voltage and indicate that as the number of GFM inverters increases in the system, the magnitude of frequency deviation during a fault increases while the duration of post-fault oscillations decreases. This is mainly because inverters have essentially very small inertial momentum but very high damping capability, which leads to faster dissipation of oscillations. The voltage deviations are similar in all scenarios but as the number of GFMs grow, the post-fault voltage transients decrease, mainly because both SG and GFM technologies have small voltage transient time constants, but GFM enjoys a much greater damping capability. These results are also evidence that greater homogeneity in the system could improve the stability; the first and fourth scenarios exhibit greater synchronization, including during the extreme transients. These results show the capability of our plane wave model to capture modern power system dynamics with a preponderance of IBRs. In the next section, we study the application of this model and shift our attention towards plane wave stability.

	\section{Discussion}
	
	\subsection*{Plane Wave Stability in Modern Electric Power Networks}

	Conventional synchronous generators have a dominantly electromechanical mechanism for frequency response, and an electromagnetic mechanism for voltage regulation. This allowed their treatment as independent problems because of the orders-of-magnitude timescale separation. However, IBRs have a fundamentally electromagnetic mechanism for the regulation of frequency and voltage. This means that in the emerging power grids with up to 100\% inverter-based renewables voltage and frequency stability need to be treated simultaneously and on the same timescale. We argue that simultaneous treatment of dynamic variables can be categorically recognized as plane wave stability problems. Events that do not result in topological changes in the system are small-signal plane wave stability problems, e.g., modal analysis or the forced oscillation problem, whereas those that result in topological changes, whether permanent or temporarily, are large-signal plane wave stability problems, e.g., the grid strength problem.
	
	To dissect each stability category, we leverage the plane wave dynamic equation for a power system that consists of a generator at node $g$, that supplies a load at node $l$, through a transformer and transmission line. In this system, the dynamics of frequency and voltage magnitude at the generator node $\dot{\omega}_g$ and $\dot{V_g}$ can be described as:
	\begin{align}
		\begin{split}  \label{eq:planar_eq}
			\begin{bmatrix}
				\dot{\omega}_g \\ \dot{V_g}
			\end{bmatrix} 
			=\underbrace{\begin{bmatrix}
					\frac{1}{\mathbf{M}_g}   & 0 \\
					0 & \frac{1}{T_{v_g}}
			\end{bmatrix}}_{\boldsymbol{\phi_n}}
			\cdot
			\begin{bmatrix}
				P_g -P_l    \\   E_g - E_l
			\end{bmatrix}
			-
			\underbrace{\begin{bmatrix}
					\frac{1}{\mathbf{M}_g}   & 0 \\
					0 & \frac{1}{T_{v_g}}
			\end{bmatrix}}_{\boldsymbol{\phi_n}}
			\cdot
			\begin{bmatrix}
				D_g \cdot \dot{\delta}_g   \\  | Z_m  \cdot  I_{gl} | 
			\end{bmatrix}
			+
			\underbrace{\begin{bmatrix}
					- \frac{M_l}{\mathbf{M}_g} & 0 \\
					0 &  - \frac{ T_{v_l}}{T_{v_g}}
			\end{bmatrix}}_{\boldsymbol{k_n}}
			\cdot
			\begin{bmatrix}
				\dot{\omega}_l \\   \dot{V_l}
			\end{bmatrix}
		\end{split}
	\end{align} 
	where $\dot{\omega}_l $ and $\dot{V_l}$ dynamics are the frequency and voltage at the load node, and $\omega_i=\dot{\delta_i}$. $D_g$ is the generator's damping function. $P_{g}$ and $E_{g}$ are the active power and the excitation voltage of the generator, respectively, and $P_l$ and $E_l$ are the active power consumption at the load node and the voltage delivered the terminals of the body where work is being done, respectively, respectively. $\mathbf{M}_g$ is the frequency momenta of the generator and transmission network combined and and $T_{v_g}$ is the generator's voltage transient time constant, and those of the load are $M_l$ and $T_{v_l}$. $Z_m$ is the magnetization impedance present in the circuit which is equivalent to that of generator and load combined, and $\dot{\delta}_g=\omega_g-\omega_s$ where $\omega_s$ is the synchronization angular frequency. Finally, $I_{gl}$ is the current that flows between the generator and the load.
	
	This formalism describes the generator's transient trajectory when subjected to an external disturbance and suggests that there are three distinct contributing components to its oscillations. 
	The first term, $\boldsymbol{\phi_n} \cdot \begin{bmatrix} P_g - P_l  & E_g - E_l \end{bmatrix}^T$, represents the dynamics that are excited by a power and voltage imbalance, e.g., load switch, change of network topology, and are contributed by available generation capacity (headroom reserve) to meet the consumption of load and the magnetization of the network. 
	The second term, $\boldsymbol{\phi_n} \cdot \begin{bmatrix}D_g \cdot \dot{\delta}_g   & | Z_m  \cdot  I_{gl}|  \end{bmatrix}^T$, represents the damping functions that describe the generator's ability to dissipate the transients and return to an equilibrium. Whilst $D_g$ serve as the frequency damping function, $Z_m$ acts as the voltage damping coefficient. 
	The third term in this equation on the right hand side, $\boldsymbol{k_n} \cdot \begin{bmatrix} \dot{\omega}_l &  \dot{V_l} \end{bmatrix}^T$, corresponds to the interactive dynamics that are excited in the adjacent nodes subsequent to the power imbalance or a load-driven disturbance that produces dynamic motions. 
	The arrays of  $\boldsymbol{\Phi_n}$, are reciprocal values of the frequency momenta of generator and line combined and the voltage transient time constant of generator.
	The arrays of $\boldsymbol{k_n}$ are negative values of the ratio of the frequency momentum and voltage transient time constant of load, to those of the network. The negative values imply different direction of oscillations. The third term becomes relevant only when the system is serving loads with high inertia, e.g., large motors, mainly industrial three phase motors, presenting a special case. Henceforth, we restrict our attention to the characteristics of the generator and the transmission line and ignore the load dynamics.
	
	Throughout the the remainder of this section, we demonstrate the numerical results of parametric sensitivity analysis of plane wave dynamic stability for the network and generators in support of Eq. \ref{eq:planar_eq}. We use the 3-generator network with two GFM and one SG from the previous section - Case (c) - with a particular interest in this configuration, as it has the lowest inertia amongst the cases that involve interactive dynamics of IBRs and SG. All simulations in this sections are subjected to a symmetrical three-phase short-circuit fault that was cleared after, $8.3$ milliseconds, as an extreme transient; thus the findings here should also hold for a small-signal event. In each scenario, one parameter was changed. The plots depict the response of generator 2. The results from the other generators were similar, thus not shown.

	\subsection*{Small-Signal Plane Wave Stability}
	
	The small-signal plane wave stability problem, commonly presented as a problem of modal analysis - both natural and force oscillations - remains an open inquiry with the increased observation of forced oscillations around the world, e.g., in the North American power grid \cite{vanfretti2011application} and in the Australian power grid - Southern Australia (West Murray Zone) \cite{mayer2023improving}. For such small-signal stability problems, if there is any power imbalance it tends to remain small and can be easily resolved. For such analysis, the control systems are required to be explicitly modelled in order to map out their interactions and identify sources of undesired oscillations. While modal analysis is well established in the literature and a mature set of tools exist (eigenvalue analysis), we find the problem of forced oscillations in power systems that have been on the rise very interesting. 
	
\begin{figure}[h]
		\centering
		\begin{subfigure}[b]{.32\textwidth}
			\includegraphics[width=1\textwidth]{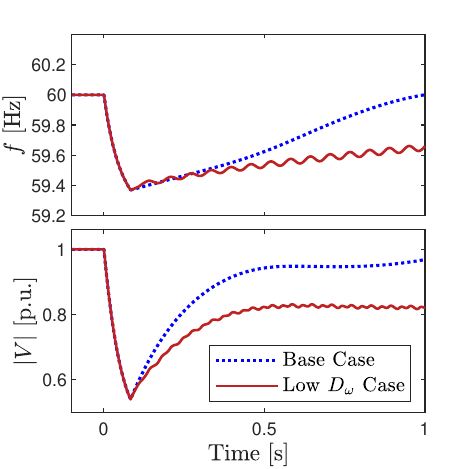}  
			\caption{~}
			\label{fig:fig2_case_low_Dw}
		\end{subfigure}  
		\begin{subfigure}[b]{.32\textwidth}     
			\centering
			\includegraphics[width=1\textwidth]{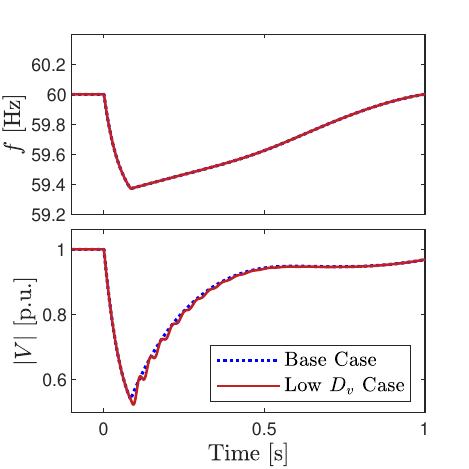}  
			\caption{~}
			\label{fig:fig2_case_low_Dv} 
		\end{subfigure} 
		\caption{\small{Parametric sensitivity of plane wave to small-signal stability; controller malfunction and forced oscillations. $D_\omega$ and $D_v$ represent corrupt frequency and voltage controllers, respectively, that induce oscillations in electric angle and magnetic flux linkage, respectively. The results indicate the frequency oscillations can propagate into voltage oscillations, evident in (\ref{fig:fig2_case_low_Dw}), but the voltage oscillations may not infect the frequency dynamics, thus the voltage oscillations remain locally bounded, evident in  (\ref{fig:fig2_case_low_Dv}).
				%
		}}
		\label{fig:forced}
	\end{figure}

	The forced oscillations, and more broadly small-signal stability problems in power systems, are fundamentally rooted in a fascinating distinction between the excitement of frequency oscillations instead of voltage oscillations. Frequency oscillations are naturally excited in the presence of any electric angle difference or frequency difference between the two adjacent nodes, whether sustained or transient. This is mainly because the intensity of the magnetic field around a transmission line is determined by the difference of electric angles at its two ends and therefore for a stable synchronization, the electric angles need to evolve at the same rate \cite{sajadi2022synchronization}. However, a reasonable difference in voltage magnitude between two adjacent nodes does not result in oscillations because they are local components, unless the difference is excessively large. This rarely occurs in practice as a transmission line would need to reach an inoperable point for this condition to materialize. Consequently, unlike frequency oscillations, voltage oscillations do not occur naturally and are always either driven by frequency oscillations or caused by the poor performance of a voltage controller or a malfunction that produces cyclical reactive power output that induce oscillations in the formation of magnetic flux. Conventionally, power system stabilizers (PSS) helped with the prevention of such oscillations greatly. If a corrupt controller produces voltage oscillations its impact will be locally bounded, whereas frequency oscillations can easily propagate. The characteristics of IBRs have increased the occurrence of forced oscillations due to the more direct involvement of digital controllers - especially frequency controllers - and the lack of mechanical components. The numerical results presented in Fig. \ref{fig:forced} support this hypothesis. Our characterization is consistent with the concerns about IBR control-driven instabilities and the deterioration of stabilizer capability that has been established in the literature \cite{matevosyan2021future}. 
	
	\subsection*{Large-Signal Plane Wave Stability}
	
	The large-signal plane wave stability problem, commonly known as power grid strength, if the resultant oscillations do not interrupt power delivery, and transient stability if an interruption occurs, continues to be a major challenge for utilities around the world and a relatively open problem in the power systems literature. Challenges have recently been observed in both the North American power grid \cite{nerc_forced,WECC_ufls} and in the Australian power grid, e.g. North Western Victoria and South Western New South Wales \cite{geddes2019power}. Because of the structural change in the system during large-signal disturbances that result in extreme transients, often a significant power imbalance becomes present. Accordingly, we identify three prominent contributing factors to the strength and the transient stability of a power network: (i) the generators' ability to supply sufficient power during transient operation, (ii) the generator's dynamic characteristics; and (iii) the transmission network's characteristics.

	\begin{figure}[!b]
		\centering
		\begin{subfigure}[b]{.32\textwidth}     
			\centering
			\includegraphics[width=1\textwidth]{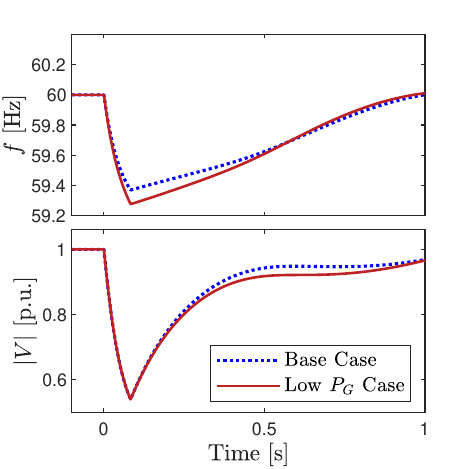}  
			\caption{~}
			\label{fig:fig2_case_low_Pg} 
		\end{subfigure}  
		\begin{subfigure}[b]{.32\textwidth}     
			\centering
			\includegraphics[width=1\textwidth]{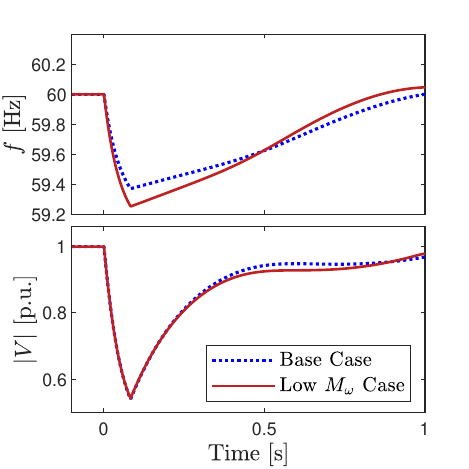}  
			\caption{~}
			\label{fig:fig2_case_low_Mw} 
		\end{subfigure} 
		\begin{subfigure}[b]{.32\textwidth}
			\includegraphics[width=1\textwidth]{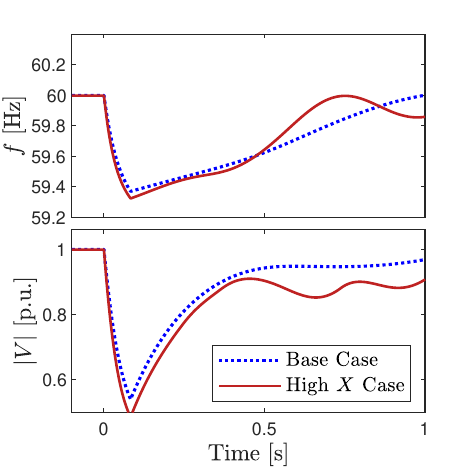} 
			\caption{~}
			\label{fig:fig2_case_high_X}
		\end{subfigure}     
		\begin{subfigure}[b]{.32\textwidth}     
			\centering
			\includegraphics[width=1\textwidth]{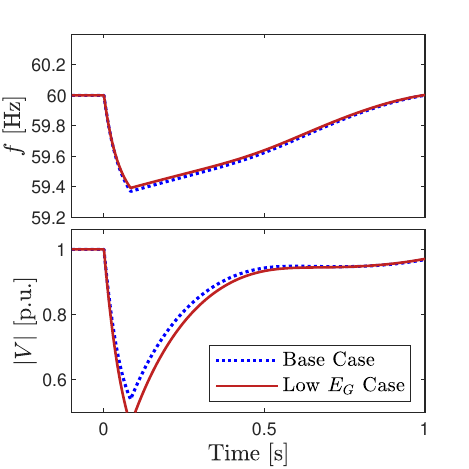}  
			\caption{~}
			\label{fig:fig2_case_low_Qf} 
		\end{subfigure} 
		\begin{subfigure}[b]{.32\textwidth}     
			\centering
			\includegraphics[width=1\textwidth]{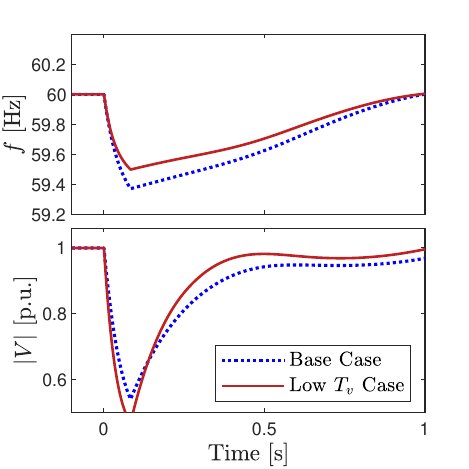}  
			\caption{~}
			\label{fig:fig2_case_low_Mv} 
		\end{subfigure} 
		\begin{subfigure}[b]{.32\textwidth}   
			\includegraphics[width=1\textwidth]{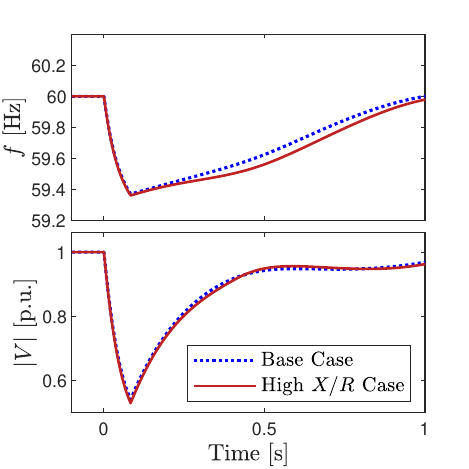} 	
			\caption{~}
			\label{fig:fig2_case_high_XR} 
		\end{subfigure}   
		\caption{\small{Parametric sensitivity to large-signal plane wave stability; grid strength. The results indicate the significance of three contributing factors to the grid stability: (i) the ability of generators to supply sufficient power during transient operation - (\ref{fig:fig2_case_low_Pg}) and (\ref{fig:fig2_case_low_Qf}), (ii) the generator's dynamic characteristics - (\ref{fig:fig2_case_low_Mw}) and (\ref{fig:fig2_case_low_Mv}), (iii) the transmission network's characteristics - (\ref{fig:fig2_case_high_X}) and (\ref{fig:fig2_case_high_XR}).
				%
		}}
		\label{fig:strength}
	\end{figure}
	
	The first contributing factor relates to the ability of the generators to supply active power to prevent extreme frequency and reactive to sustain the excitation voltage during transients and prevent excessive voltage drops, evident in Fig. (\ref{fig:fig2_case_low_Pg}) and Fig.  (\ref{fig:fig2_case_low_Qf}). The analogous metric used to describe this factor is the short circuit ratio (SCR) of generators. The SCR ratio has been utilized and proven practical as a heuristic solution where a higher SCR implies a stronger grid. It has been used as a rule of thumb approach and a practical proxy to estimate the availability of synchronous generator-like support during extreme transients \cite{dozein2018system} without a well-vetted scientific explanation. Here, we scientifically establish the use of SCR as a reasonable measure as it quantifies how well a power network can maintain the magnetic field, and subsequently the electric field. Obviously, a high SCR is practical only if robust electromagnetic properties for the interconnecting transmission line are provided so that it is capable of sustaining the electromagnetic field even during extreme transients. The SCR is more broadly addressed in the context of the $P-Q$ capability characteristics of a generator that determines its ability to absorb or inject active and reactive power \cite{dozein2018system} ($P$ and $Q$ are active and reactive power, respectively), whereas the transmission line capability is discussed in the context of the network transfer capability, especially in dealing with thermal stability and voltage stability. 
	The generators with constrained capability to provide active and reactive power support during an extreme transient event, e.g. IBRs during an electrical short-circuit since the limiting factor is the use of semiconductor switches, could comprise grid stability and strength. 
	This explains why synchronous condenser technologies have been widely recognized as a solution to the grid strength problem \cite{trujillo2022operability,kenyon2020grid,hadavi2021robust}, simply because the high short circuit current capability they offer prevents demagnetization of the line through transient reactive power to support the electromagnetic field, in addition to the inertial support they provide for a smoother frequency response. Reactive power support from a synchronous condenser during a fault and after its clearance also relieves IBRs with GFM control from some of their reactive power (voltage) support, allowing them to provide active power during the fault and preventing a significant frequency drop with more effective frequency damping, which also serves to dampen voltage oscillations. 
	
	The second contributing factor is the generator's dynamic characteristics; that are the inertial frequency momentum and the voltage transient time constant, as shown in Figs. (\ref{fig:fig2_case_low_Mw}) and  (\ref{fig:fig2_case_low_Mv}). While it is evident that the reduced generator inertia and voltage transient time constant will result in more deviations and frequency and voltage transients, it also results in a faster recovery subsequent to the clearance of the external disturbance. Hence, we suggest high shares of IBR could result in larger frequency and voltage deviation during a disturbance, but if there are IBR GFM resources present, because of their enhanced damping capability and low inertia, both frequency and voltage oscillations will be mitigated rapidly. Mass integration of GFL technology alone can pose stability challenges because of the lack of damping capability, as evidenced by the majority of the instabilities witnessed in the real-world, e.g., recent blackouts in North America: the interruption in California on July 7, 2020 \cite{sanfernando20} and the interruptions in Texas on May 9, 2021, June 26, 2021 \cite{odessa21}, and June 4, 2022 \cite{odessa22}.

	The third contributing factor relates to the network characteristics. We identify that networks with higher $X$ values are more susceptible to more extreme oscillations - as shown in Fig. (\textbf{\ref{fig:fig2_case_high_X}}). Two distinct root causes can be considered to explain this phenomenon. First, the damping ratio of oscillations in the system, $\zeta$, is inversely proportional to the $X$ value. In effect, higher $X$ values reduce the damping capability of the network. Second, transferring a given amount of active power in a circuit that has a larger $X$ reactance while maintaining unitary voltages causes the transfer angle, $\delta$ to increase. Since $P=\frac{1}{X} sin \delta$, therefore $sin \delta$ increases proportionally to the higher $X$ value, and consequently $cos \delta$ decreases. This means that to transfer a fixed amount of active power in a network with higher $X$ values, the reactive power transfer is constrained to a lower amount, $Q=\frac{1}{X}(cos \delta - 1)$. As a result, the voltage is more severely impacted by a fault, not because of the change in voltage transient time constant - but because the reactive power supply that can maintain the voltage is smaller, 
	thus the voltage drops to a lower point. The above-mentioned root causes suggest higher values of $X$ yield more severe frequency and voltage oscillations and, if sufficient reactive power headroom is not available when trying to sustain voltage during transients (a common drawback issue with GFL IBRs), the voltage drops are greater, as shown in Fig. \ref{fig:fig2_case_high_X}. Finally, while the $\frac{X}{R}$ ratio is widely used by power engineers to quantify grid strength \cite{adib2018stability,dozein2022dynamics, badrzadeh2021system,yang2014impedance}, where $X$ and $R$ are the inductance and resistance of the lines, respectively, we do not find reasonable grounds to consider the $\frac{X}{R}$ ratio as a meaningful or effective means to quantify the grid strength in high voltage transmission networks, as evident from Fig. (\ref{fig:fig2_case_high_XR}). We recognize that this metric might be effective in low voltage networks, e.g., microgrids or distribution networks, but as the voltage rating increases, the $R$ value per mile becomes smaller while the changes in $X$ value per mile are quite small (reactance changes with the logarithm of conductor radius) \cite{short2003electric}, which is why in analysis of high voltage transmission networks, the ohmic losses are often neglected. Therefore, in high voltage networks, the $\frac{X}{R}$ loses its practical meaning.

	\section{Outlook and Closing Remarks}
	
	The world is warming at an unprecedented and accelerated rate and a key impediment of cross-sectoral decarbonization efforts is currently achieving the ability to operate power grids with ultra-high shares of variable renewable energy \cite{kroposki2017achieving} as the cost of renewable energy generation continues to decline \cite{Lazard2023}. Our plane wave dynamic model of electric power networks highlighted the significance of the electromagnetic phenomena in power grids that is mainly overlooked in the contemporary models. It also linked the frequency and voltage dynamics, as they are directly linked in an electromagnetic field, consistent with the fundamental physical nature of the electricity. Subsequently, the two dimensional inference of stability in power grids established the contributing factors to the recently observed phenomena in power systems around the world that the conventional dynamic models could not definitively explain, including forced oscillations and grid strength.

	This research has implications for the broader scientific community and direct applications that will benefit society at large. The consensus amongst scholars and practitioners is that achieving a 100\% decarbonized power grid that has the capability to operate with ultra-high shares of renewables is currently hindered by the open question of how to replace the abundance of inertia and synchronous torque that the synchronous generators in fossil-fuel and nuclear power plants currently provide \cite{kroposki2017achieving}. Our model estimates that the momentum that is stored in the electromagnetic field around the transmission network in the United States, that has been thus far unaccounted for in power system dynamics, if partially loaded nationwide (50\% of the carrying capacity), could provide as much momentum as $51$ medium-sized generating units that are typically housed in thermal or nuclear power plants with an inertial constant of $H=6s$ at $500$ MVA rating offer. This would be roughly equivalent to $6$ in the Texas Interconnection (ERCOT), $30$ in the Eastern Interconnection, and $15$ in the Western Interconnection - based on the scale of each interconnection. In practice, the length of many of the lines is shorter than the maximum viable transfer distance of their voltage class and they are not always loaded at their carrying capacity, therefore, the quantity of electromagnetic momentum may vary from our estimation. Nevertheless, knowing the power system in the United States nationwide has a few thousand synchronous generators, this estimation hints at the energy that may be available in the electromagnetic field around the line that is presently not considered. After all, this is only a fraction of the momentum provided by the mechanical machines in thermal and hydro power plants. Assuming the share of electromagnetic momentum is only $10-15\%$ of the overall momentum stored in the system (according to the results shown in Figs. \ref{fig:9-Ms} and \ref{fig:39-Ms}), then consequently, it means the transmission network alone in the United States could account for an overall inertial constant in the neighborhood of $H \approx 0.5-1.0$s. This is currently present and embodied in the system, and is a contributing component to the inertial constant of $H \approx 5-6.5$s measured in the system \cite{tucker2022frequency}. Even though this number may seem small, it is still much greater than what IBRs can provide, and thus opens up the possibility of all-IBR systems. From the operational perspective, the electromagnetic momentum of the lines could become significant during operating conditions with high shares of IBR and may serve as the critical inertia needed to slow down the frequency/voltage dynamics just enough to deploy the fast frequency responsive resources that provide power balancing and damping (particularly from GFM inverters) before the system-wide frequency drops below its operational limit. This assumes that transient support can be provided during short-circuit faults. From the planning perspective, the findings of this study can support forward-looking long-range grid planning processes by identifying the resources and characteristics needed to ensure the resilience and security in 20-30 years when significantly larger contributions from solar, wind, and storage units are expected than today.

	We believe our developments in this are groundbreaking as they unveil a large amount of electromagnetic momentum present in the system that can be utilized to pave the path towards a fully decarbonized electric power sector. They also set the scientific basis for the development of control solutions for the reliable, safe, and stable operation of future all-renewable power networks. How generators are controlled determines the reliability and stability of power delivery by defining how the electromagnetic link between the generators and the loads is kept stable and robust. We believe that in future power system operations with significantly high shares of IBRs we will move from frequency and voltage control in generators towards incorporation of power factor control in generators. Frequency and voltage control determines the magnitude of electric and magnetic fields without much control over their phase, while power factor control is directly synonymous to control of how electromagnetic waves are polarized by controlling both the amplitude and the phase of magnetic and electric fields.

	
	\subsection*{Data Availability}
	The PSCAD models used to produce results presented in Fig. 1 are available open-source on Github via https://github.com/NREL/PyPSCAD.
	The raw data to generate Fig. 2 is restricted for public access. 
	The data to reproduce the results presented in Fig. 3, Fig. 4, and Fig. 5 is provided throughout this paper.


	\section*{Acknowledgments}
	
	We wish to thank I.~Dobson of Iowa State University, R.~W.~Kenyon and M.~Trujillo of University of Colorado Boulder, and B.~Kroposki and A.~Hoke of the National Renewable Energy Laboratory (NREL), for their insightful discussions and critical comments and H.~Scribner and M.~Nugent of the Southwest Power Pool (SPP) for providing us with the high frequency measurement data of the SPP power network.

	This work was authored in part by the National Renewable Energy Laboratory, operated by Alliance for Sustainable Energy, LLC, for the U.S. Department of Energy (DOE). The views expressed in the article do not necessarily represent the views of the DOE or the U.S. Government. The U.S. Government retains and the publisher, by accepting the article for publication, acknowledges that the U.S. Government retains a nonexclusive, paid-up, irrevocable, worldwide license to publish or reproduce the published form of this work, or allow others to do so, for U.S. 
	
	This work was supported by the U.S. Department of Energy under Contract No. DE-AC36-08-GO28308 with the National Renewable Energy Laboratory.
	
	\section{Author Contributions Statement}
	
	The conceptual approach and research design were performed by A.~Sajadi with the analytical and numerical analyses carried out by A.~Sajadi. B.~M.~Hodge aided in the conceptual approach and supervised the project. All authors contributed to writing the manuscript.
	
	\section{Funding}
	Authors have no funding to report. 
	
	\section{Competing Interests Statement}
	Authors have no competing interest to report.



\begin{thebibliography}{10}
		
		\bibitem{motter2013spontaneous}
		A.~E. Motter, S.~A. Myers, M.~Anghel, and T.~Nishikawa, ``Spontaneous synchrony
		in power-grid networks,'' {\em Nature Physics}, vol.~9, no.~3, pp.~191--197,
		2013.
		
		\bibitem{denholm2020inertia}
		P.~Denholm, T.~Mai, R.~W. Kenyon, B.~Kroposki, and M.~O'Malley, ``Inertia and
		the power grid: A guide without the spin,'' tech. rep., National Renewable
		Energy Lab.(NREL), Golden, CO (United States), 2020.
		
		\bibitem{liacco1967adaptive}
		T.~E.~D. Liacco, ``The adaptive reliability control system,'' {\em IEEE
			Transactions on Power Apparatus and Systems}, no.~5, pp.~517--531, 1967.
		
		\bibitem{ye2016analysis}
		H.~Ye, Y.~Liu, P.~Zhang, and Z.~Du, ``Analysis and detection of forced
		oscillation in power system,'' {\em IEEE Transactions on Power Systems},
		vol.~32, no.~2, pp.~1149--1160, 2016.
		
		\bibitem{sarmadi2015inter}
		S.~A.~N. Sarmadi and V.~Venkatasubramanian, ``Inter-area resonance in power
		systems from forced oscillations,'' {\em IEEE Transactions on Power Systems},
		vol.~31, no.~1, pp.~378--386, 2015.
		
		\bibitem{dozein2021simultaneous}
		M.~G. Dozein, O.~Gomis-Bellmunt, and P.~Mancarella, ``Simultaneous provision of
		dynamic active and reactive power response from utility-scale battery energy
		storage systems in weak grids,'' {\em IEEE Transactions on Power Systems},
		vol.~36, no.~6, pp.~5548--5557, 2021.
		
		\bibitem{AEMO2020strength}
		``System strength in the nem explained,'' tech. rep., Australian Energy Market
		Operator (AEMO), 2020.
		
		\bibitem{machowski2020power}
		J.~Machowski, Z.~Lubosny, J.~W. Bialek, and J.~R. Bumby, {\em Power system
			dynamics: stability and control}.
		\newblock John Wiley \& Sons, 2020.
		
		\bibitem{milano2021complex}
		F.~Milano, ``Complex frequency,'' {\em IEEE Transactions on Power Systems},
		vol.~37, no.~2, pp.~1230--1240, 2021.
		
		\bibitem{milano2021geometrical}
		F.~Milano, ``A geometrical interpretation of frequency,'' {\em IEEE
			Transactions on Power Systems}, vol.~37, no.~1, pp.~816--819, 2021.
		
		\bibitem{moutevelis2023taxonomy}
		D.~Moutevelis, J.~Rold{\'a}n-P{\'e}rez, M.~Prodanovic, and F.~Milano,
		``Taxonomy of power converter control schemes based on the complex frequency
		concept,'' {\em IEEE Transactions on Power Systems}, 2023.
		
		\bibitem{he2022complex}
		X.~He, V.~H{\"a}berle, and F.~D{\"o}rfler, ``Complex-frequency synchronization
		of converter-based power systems,'' {\em arXiv preprint arXiv:2208.13860},
		2022.
		
		\bibitem{yang2021augmented}
		P.~Yang, F.~Liu, T.~Liu, and D.~J. Hill, ``Augmented synchronization of power
		systems,'' {\em arXiv preprint arXiv:2106.13166}, 2021.
		
		\bibitem{li2023intrinsic}
		Y.~Li, T.~C. Green, and Y.~Gu, ``The intrinsic communication in power systems:
		A new perspective to understand synchronization stability,'' {\em IEEE
			Transactions on Circuits and Systems I: Regular Papers}, 2023.
		
		\bibitem{olsen2015high}
		R.~G. Olsen, {\em High Voltage Overhead Transmission Line Electromagnetics}.
		\newblock CreateSpace, an Amazon Company, 2015.
		
		\bibitem{emanuel2011power}
		A.~E. Emanuel, {\em Power definitions and the physical mechanism of power
			flow}, vol.~22.
		\newblock John Wiley \& Sons, 2011.
		
		\bibitem{kirtley2020electric}
		J.~L. Kirtley, {\em Electric power principles: sources, conversion,
			distribution and use}.
		\newblock John Wiley \& Sons, 2020.
		
		\bibitem{bergen2009power}
		A.~R. Bergen, {\em Power systems analysis}.
		\newblock Pearson Education India, 2009.
		
		\bibitem{sauer2017power}
		P.~W. Sauer, M.~A. Pai, and J.~H. Chow, {\em Power system dynamics and
			stability: with synchrophasor measurement and power system toolbox}.
		\newblock John Wiley \& Sons, 2017.
		
		\bibitem{poynting1884xv}
		J.~H. Poynting, ``Xv. on the transfer of energy in the electromagnetic field,''
		{\em Philosophical Transactions of the Royal Society of London}, no.~175,
		pp.~343--361, 1884.
		
		\bibitem{maxwell1865viii}
		J.~C. Maxwell, ``Viii. a dynamical theory of the electromagnetic field,'' {\em
			Philosophical transactions of the Royal Society of London}, no.~155,
		pp.~459--512, 1865.
		
		\bibitem{collin1990field}
		R.~E. Collin, {\em Field theory of guided waves}, vol.~5.
		\newblock John Wiley \& Sons, 1990.
		
		\bibitem{emanuel2004poynting}
		A.~E. Emanuel, ``Poynting vector and the physical meaning of nonactive
		powers,'' {\em IEEE Transactions on Instrumentation and Measurement},
		vol.~54, no.~4, pp.~1457--1462, 2005.
		
		\bibitem{balanis2012advanced}
		C.~A. Balanis, {\em Advanced engineering electromagnetics}.
		\newblock John Wiley \& Sons, 2012.
		
		\bibitem{wolski2011theory}
		A.~Wolski, ``Theory of electromagnetic fields,'' {\em arXiv preprint
			arXiv:1111.4354}, 2011.
		
		\bibitem{thomson284elements}
		J.~Thomson, ``Elements of the mathematical theory of electricity and magnetism
		(cambridge, london, 1904).''
		
		\bibitem{johnson1994electromagnetic}
		F.~S. Johnson, B.~L. Cragin, and R.~R. Hodges, ``Electromagnetic momentum
		density and the poynting vector in static fields,'' {\em American journal of
			physics}, vol.~62, no.~1, pp.~33--41, 1994.
		
		\bibitem{bak1994energy}
		D.~Bak, D.~Cangemi, and R.~Jackiw, ``Energy-momentum conservation in gravity
		theories,'' {\em Physical Review D}, vol.~49, no.~10, p.~5173, 1994.
		
		\bibitem{beck2007virtual}
		H.-P. Beck and R.~Hesse, ``Virtual synchronous machine,'' in {\em 2007 9th
			international conference on electrical power quality and utilisation},
		pp.~1--6, IEEE, 2007.
		
		\bibitem{lasseter2019grid}
		R.~H. Lasseter, Z.~Chen, and D.~Pattabiraman, ``Grid-forming inverters: A
		critical asset for the power grid,'' {\em IEEE Journal of Emerging and
			Selected Topics in Power Electronics}, vol.~8, no.~2, pp.~925--935, 2019.
		
		\bibitem{sinha2015uncovering}
		M.~Sinha, F.~D{\"o}rfler, B.~B. Johnson, and S.~V. Dhople, ``Uncovering droop
		control laws embedded within the nonlinear dynamics of van der pol
		oscillators,'' {\em IEEE Transactions on Control of Network Systems}, vol.~4,
		no.~2, pp.~347--358, 2015.
		
		\bibitem{athay1979practical}
		T.~Athay, R.~Podmore, and S.~Virmani, ``A practical method for the direct
		analysis of transient stability,'' {\em IEEE Transactions on Power Apparatus
			and Systems}, no.~2, pp.~573--584, 1979.
		
		\bibitem{fernandez2018coding}
		A.~Fern{\'a}ndez~Rodr{\'\i}guez, L.~de~Santiago~Rodrigo,
		E.~L{\'o}pez~Guill{\'e}n, J.~M. Rodr{\'\i}guez~Ascariz, J.~M.
		Miguel~Jim{\'e}nez, and L.~Boquete, ``Coding prony's method in matlab and
		applying it to biomedical signal filtering,'' {\em BMC bioinformatics},
		vol.~19, pp.~1--14, 2018.
		
		\bibitem{sajadi2022synchronization}
		A.~Sajadi, R.~W. Kenyon, and B.-M. Hodge, ``Synchronization in electric power
		networks with inherent heterogeneity up to 100\% inverter-based renewable
		generation,'' {\em Nature communications}, vol.~13, no.~1, pp.~1--12, 2022.
		
		\bibitem{NERC_UFLS_standard}
		{North American Electric Reliability Corporation (NERC)}, ``Prc-006-npcc-2 -
		automatic underfrequency load shedding.,'' 2021.
		
		\bibitem{vanfretti2011application}
		L.~Vanfretti, L.~Dosiek, J.~W. Pierre, D.~Trudnowski, J.~H. Chow,
		R.~Garc{\'\i}a-Valle, and U.~Aliyu, ``Application of ambient analysis
		techniques for the estimation of electromechanical oscillations from measured
		pmu data in four different power systems,'' {\em European Transactions on
			Electrical Power}, vol.~21, no.~4, pp.~1640--1656, 2011.
		
		\bibitem{mayer2023improving}
		P.~F. Mayer, M.~Gordon, W.-C. Huang, and C.~Hardt, ``Improving grid strength in
		a wide-area transmission system with grid forming inverters,'' {\em IET
			Generation, Transmission \& Distribution}, vol.~17, no.~2, pp.~399--410,
		2023.
		
		\bibitem{matevosyan2021future}
		J.~Matevosyan, J.~MacDowell, N.~Miller, B.~Badrzadeh, D.~Ramasubramanian,
		A.~Isaacs, R.~Quint, E.~Quitmann, R.~Pfeiffer, H.~Urdal, {\em et~al.}, ``A
		future with inverter-based resources: Finding strength from traditional
		weakness,'' {\em IEEE Power and Energy Magazine}, vol.~19, no.~6, pp.~18--28,
		2021.
		
		\bibitem{nerc_forced}
		{North America Electric Reliability Corporation (NERC)}, ``Eastern
		interconnection oscillation disturbance: January 11, 2019 forced oscillation
		event,'' 2019.
		
		\bibitem{WECC_ufls}
		{Western Electricity Coordinating Council (WECC)}, ``2021 ufls assessment,''
		2022.
		
		\bibitem{geddes2019power}
		J.~Geddes, ``Power system limitations in north western victoria and south
		western new south wales,'' 2019.
		
		\bibitem{dozein2018system}
		M.~G. Dozein, P.~Mancarella, T.~K. Saha, and R.~Yan, ``System strength and weak
		grids: Fundamentals, challenges, and mitigation strategies,'' in {\em 2018
			Australasian Universities Power Engineering Conference (AUPEC)}, pp.~1--7,
		IEEE, 2018.
		
		\bibitem{trujillo2022operability}
		M.~Trujillo, R.~W. Kenyon, G.~Yau, L.~Yu, A.~Hoke, and B.-M. Hodge,
		``Operability of a power system with synchronous condensers and
		grid-following inverters,'' in {\em 2022 IEEE 49th Photovoltaics Specialists
			Conference (PVSC)}, pp.~1038--1042, IEEE, 2022.
		
		\bibitem{kenyon2020grid}
		R.~W. Kenyon, A.~Hoke, J.~Tan, and B.-M. Hodge, ``Grid-following inverters and
		synchronous condensers: A grid-forming pair?,'' in {\em 2020 Clemson
			University Power Systems Conference (PSC)}, pp.~1--7, IEEE, 2020.
		
		\bibitem{hadavi2021robust}
		S.~Hadavi, D.~B. Rathnayake, G.~Jayasinghe, A.~Mehrizi-Sani, and B.~Bahrani,
		``A robust exciter controller design for synchronous condensers in weak
		grids,'' {\em IEEE Transactions on Power Systems}, vol.~37, no.~3,
		pp.~1857--1867, 2021.
		
		\bibitem{sanfernando20}
		{Joint NERC and Texas RE Staff}, ``San fernando disturbance, southern
		california event: July 7, 2020,'' {\em NERC, Atlanta, GA, USA}, 2020.
		
		\bibitem{odessa21}
		{Joint NERC and Texas RE Staff}, ``Odessa disturbance texas events: May 9, 2021
		and june 26, 2021,'' {\em NERC, Atlanta, GA, USA}, 2021.
		
		\bibitem{odessa22}
		{Joint NERC and Texas RE Staff}, ``Odessa disturbance texas event: June 4,
		2022,'' {\em NERC, Atlanta, GA, USA}, 2022.
		
		\bibitem{adib2018stability}
		A.~Adib, B.~Mirafzal, X.~Wang, and F.~Blaabjerg, ``On stability of voltage
		source inverters in weak grids,'' {\em Ieee Access}, vol.~6, pp.~4427--4439,
		2018.
		
		\bibitem{dozein2022dynamics}
		M.~G. Dozein, B.~C. Pal, and P.~Mancarella, ``Dynamics of inverter-based
		resources in weak distribution grids,'' {\em IEEE Transactions on Power
			Systems}, vol.~37, no.~5, pp.~3682--3692, 2022.
		
		\bibitem{badrzadeh2021system}
		B.~Badrzadeh, Z.~Emin, S.~Goyal, S.~Grogan, A.~Haddadi, A.~Halley, A.~Louis,
		T.~Lund, J.~Matevosyan, T.~Morton, {\em et~al.}, ``System strength,'' {\em
			CIGRE Science\&Engineering Journal}, no.~20, 2021.
		
		\bibitem{yang2014impedance}
		D.~Yang, X.~Ruan, and H.~Wu, ``Impedance shaping of the grid-connected inverter
		with lcl filter to improve its adaptability to the weak grid condition,''
		{\em IEEE Transactions on Power Electronics}, vol.~29, no.~11,
		pp.~5795--5805, 2014.
		
		\bibitem{short2003electric}
		T.~A. Short, {\em Electric power distribution handbook}.
		\newblock CRC press, 2003.
		
		\bibitem{kroposki2017achieving}
		B.~Kroposki, B.~Johnson, Y.~Zhang, V.~Gevorgian, P.~Denholm, B.-M. Hodge, and
		B.~Hannegan, ``Achieving a 100\% renewable grid: Operating electric power
		systems with extremely high levels of variable renewable energy,'' {\em IEEE
			Power and energy magazine}, vol.~15, no.~2, pp.~61--73, 2017.
		
		\bibitem{Lazard2023}
		{Lazard Consultant}, {\em {Levelized Cost of Energy and Levelized Cost of
				Storage 2023}}, 2023 (accessed February 8, 2023).
		
		\bibitem{tucker2022frequency}
		D.~Tucker, D.~Kosterev, and A.~Sajadi, ``Frequency response in the western
		interconnection with significantly high shares of inverter-based resources:
		100\% renewable case,'' in {\em 2022 IEEE Power \& Energy Society General
			Meeting (PESGM)}, pp.~1--5, IEEE, 2022.
		
	\end{thebibliography}
\end{document}